# The Dark Energy Survey♣

T. Abbott[1], G. Aldering[2], J. Annis[3], M. Barlow[4], C. Bebek[2], B. Bigelow[5], C. Beldica[6], R. Bernstein[5], S. Bridle[4], R. Brunner[7], J. Carlstrom[8,9], M. Campbell[10], F. Castander[11], C. Cunha[8,9], H. T. Diehl[3], S. Dodelson[3,8], P. Doel[4], G. Efstathiou[12], J. Estrada[3], A. Evrard[10], E. Fernández[13], B. Flaugher[3], P. Fosalba[11], J. Frieman[3,8,9], E. Gaztañaga[11], D. Gerdes[10], M. Gladders[8,14], W. Hu[8,9], D. Huterer[8,9], B. Jain[15], I. Karliner[16], S. Kent[3,8], O. Lahav[4], M. Levi[2], M. Lima[8,9], H. Lin[3], P. Limon[3], M. Martínez[13], T. McKay[10], R. McMahon[12], K. W. Merritt[3], C. Miller[1], J. Miralda-Escude[11], J. Mohr[7,16], R. Nichol[17], H. Oyaizu[8,9], J. Peacock[18], J. Peoples[3], S. Perlmutter[2], R. Plante[6], P. Ricker[16], N. Roe[2], V. Scarpine[3], M. Schubnell[10], M. Selen[16], E. Sheldon[8,9], C. Smith[1], A. Stebbins[3], C. Stoughton[3], N. Suntzeff[1], W. Sutherland[12], M. Takada[19], G. Tarle[10], M. Tecchio[10], J. Thaler[16], D. Tucker[3], S. Viti[4], A. Walker[1], R. Wechsler[8,9], J. Weller[3,4], W. Wester[3]

[1]Cerro Tololo Inter-American Observatory, National Optical Astronomy Observatory, La Serena, Chile
[2]Lawrence Berkeley National Laboratory, Berkeley, CA, USA
[3]Fermi National Accelerator Laboratory, Batavia, IL, USA
[4]Department of Physics & Astronomy, University College, London, UK
[5]Department of Astronomy, University of Michigan, Ann Arbor, MI, USA
[6]National Center for Supercomputing Applications, University of Illinois, Urbana-Champaign, IL, USA
[7]Department of Astronomy, University of Illinois, Urbana-Champaign, IL, USA
[8]Department of Astronomy, The University of Chicago, Chicago, IL, USA
[9]Kavli Institute for Cosmological Physics, The University of Chicago, Chicago, IL, USA
[10]Department of Physics, University of Michigan, Ann Arbor, MI, USA
[11]Institut d'Estudis Espacials de Catalunya/CSIC, Barcelona, Spain
[12]Institute of Astronomy, University of Cambridge, Cambridge, UK
[13]Institut de Fisica d'Altes Energies, Barcelona, Spain
[14]Observatories of the Carnegie Institute of Washington, Pasadena, CA, USA
[15]Department of Physics & Astronomy, University of Pennsylvania, Philadelphia, PA, USA
[16]Department of Physics, University of Illinois, Urbana-Champaign, IL, USA
[17]Institute for Cosmology & Gravitation, University of Portsmouth, Portsmouth, UK
[18]Institute for Astronomy, University of Edinburgh, Edinburgh, UK
[19]Astronomical Institute, Tohoku University, Sendai, Japan

## Overview

We describe the Dark Energy Survey (DES), a proposed optical-near infrared survey of 5000 sq. deg of the South Galactic Cap to ~24$^{th}$ magnitude in SDSS *griz*, that would use a new 3 deg$^2$ CCD camera to be mounted on the Blanco 4-m telescope at Cerro Telolo Inter-American Observatory (CTIO). The survey data will allow us to measure the dark energy and dark matter densities and the dark energy equation of state through four independent methods: galaxy clusters, weak gravitational lensing tomography, galaxy angular clustering, and supernova distances. These methods are doubly complementary: they constrain different combinations of cosmological model parameters and are subject to different systematic errors. By deriving the four sets of measurements from the same data set with a common analysis framework, we will obtain important cross checks of the systematic errors and thereby make a substantial and robust advance in the precision of dark energy measurements.

---

♣ slightly modified version of a White Paper submitted to the Dark Energy Task Force, June 15, 2005.

# 1. Background

The National Optical Astronomy Observatory (NOAO) issued an announcement of opportunity (AO) in December 2003 for an open competition to partner with NOAO in building an advanced instrument for the Blanco telescope in exchange for awarding the instrument collaboration up to 30% of the observing time over a five-year period for a compelling science project. In response to this AO, the Dark Energy Survey (DES) Collaboration was formed and submitted a proposal to NOAO in July 2004 to build DECam, a new wide-field imager for the Blanco, with the goal of carrying out a survey to address the nature of the dark energy using the four primary techniques described below in Sec. 2. The Blanco Instrumentation Review Panel convened by NOAO to review the DES proposal concluded that the scientific goals are exciting and timely. Subsequently, the NOAO Director asked the CTIO Director and the DES Project Director to draft a MOU among the Parties that would define the terms of the partnership.

The major components of DECam are a 519 megapixel optical CCD camera, a wide-field optical corrector (2.2 deg. field of view), a 4-band filter system with SDSS *g, r, i,* and *z* filters, guide and focus sensors mounted on the focal plane, low-noise CCD readout, a cryogenic cooling system to maintain the focal plane at 180 K, as well as a data acquisition and instrument control system to connect to the Blanco observatory infrastructure. The camera focal plane will consist of sixty-two 2k x 4k CCDs (0.27"/pixel) arranged in a hexagon covering an imaging area of 3 sq. degrees. Smaller format CCDs for guiding and focusing will be located at the edges of the focal plane. More details about the instrument are given below in Sec. 7 and in the Supplements.

To carry out the Dark Energy Survey with DECam, we have requested 525 nights of observing on the Blanco telescope over 5 years, concentrated between September and February, beginning in Sept. 2009. With that time, we expect to reach photometric limits of *g*=24.6, *r*=24.1, *i*=24.3, and *z*=23.9 over 5000 sq. deg of sky. These are $10\sigma$ limits in 1.5" apertures assuming 0.9" seeing and are appropriate for faint galaxies; the corresponding $5\sigma$ limit for point sources is 1.5 mags fainter. These limits and adopted median delivered seeing are derived from detailed survey simulations that incorporate weather and seeing data at CTIO over a 30-year baseline.

The survey strategy is designed to optimize the photometric calibration by tiling each region of the survey with at least four overlapping pointings in each band. This provides uniformity of coverage and control of systematic photometric errors via relative photometry on scales up to the survey size. This strategy will enable us to determine photometric redshifts (photo-z's) of galaxies to an accuracy of $\sigma(z)\sim0.07$ out to z>1, with some dependence on redshift and galaxy type, cluster photometric redshifts to $\sigma(z)\sim0.02$ or better out to z~1.3, and shapes for approximately 200 million galaxies; these measurements will be sufficient to meet the survey science requirements. 4000 deg$^2$ of the survey region will overlap the South Pole Telescope Sunyaev-Zel'dovich survey region; the remainder will provide coverage of spectroscopic redshift training sets, including the SDSS southern equatorial stripe, and more complete coverage near the South Galactic pole.

This white paper is organized as follows. Section 2 briefly describes the four main techniques that DES will use, alone and in conjunction with the South Pole Telescope Sunyaev-Zel'dovich effect survey, to probe the dark energy. Section 3 presents forecast statistical constraints on the dark energy equation of state for several choices of priors on other parameters, followed by discussion of the additional assumptions that enter these forecasts. In Section 4, we discuss the primary envisioned systematic errors for each method and outline the methods of controlling them. Section 5 lists the precursor and concurrent observations and developments upon which DES will rely. In Section 6, we briefly describe DECam, the survey instrument, and Section 7 outlines the data management plans. Section 8 presents a timeline for the project, and we conclude in Section 9. The Supplement sections that follow are appendices that provide more detailed technical information on various aspects of the project.

A more comprehensive though less up-to-date description of the project is available at https://decam.fnal.gov/NOAO04/A_Proposal_to_NOAO.pdf

## 2. Dark Energy Survey Techniques

Here, we briefly summarize the four proposed techniques for probing dark energy. The forecast dark energy constraints are described in the following section. We describe these techniques and their associated uncertainties in greater detail in the **Supplements for the Dark Energy Survey**.

*Galaxy clusters*: The evolution of the galaxy cluster mass function and cluster spatial correlations provide a sensitive probe of the dark energy; these observables are affected by cosmology through both the growth of density perturbations and the evolution of the volume element (Haiman, Mohr, & Holder 2000, Battye & Weller 2003). Clusters make promising cosmological probes, because the formation of these large potential wells involves only the gravitational dynamics of dark matter to good approximation. The primary design driver of the DES is the detailed optical measurement of galaxy clusters, including photometric redshifts, in conjunction with the South Pole Telescope (SPT) Survey. The SPT (Ruhl et al 2004) will use the Sunyaev-Zel'dovich effect (SZE) to detect galaxy clusters out to large distances, providing a census of tens of thousands of clusters over a 4000 square degree region south of declination $\delta = -30^\circ$. The integrated SZE flux decrement is expected to be a robust indicator of cluster mass, because it is a measure of the total thermal energy of the electrons residing in the gravitational potential well; in particular, it should be insensitive to gas dynamics in the cluster core (Motl et al 2005, Nagai 2005). The DES is designed to measure efficiently and accurately photometric redshifts for all SPT clusters to $z=1.3$. It will also cross-check the completeness of the SPT cluster selection function by optically identifying clusters below the SPT mass threshold and will statistically calibrate SZE cluster mass estimates using the cluster-mass correlation function inferred from weak lensing (Johnston et al 2005). Existing cameras would require decades to cover the SPT survey area to the requisite depth.

*Weak lensing tomography:* The DES will measure the weak lensing (WL) shear of galaxies as a function of photometric redshift. The evolution of the statistical pattern of WL distortions—for example, the shear-shear (S-S) angular power spectrum—and of the cross-correlation between foreground galaxies and background galaxy shear (galaxy-shear correlations, G-S), are sensitive to the cosmic expansion history through both geometry and the growth rate of structure (Hu 2002, Huterer 2002). In the course of surveying 5000 sq. deg. to the depth required for cluster photo-z's, the DES will measure shapes and photometric redshifts for ~300 million galaxies and, with improved control of the optical image quality, enable accurate measurement of lensing by large-scale structure.

*Galaxy angular clustering*: The DES will measure the angular clustering of galaxies (denoted G-G in Table 1) in photometric redshift shells out to $z\sim1.1$. The matter power spectrum as a function of wave-number shows characteristic features, a broad peak as well as baryon wiggles arising from the same acoustic oscillations that give rise to the Doppler peaks in the CMB power spectrum; these features were recently detected in the SDSS (Eisenstein et al 2005). In combination with CMB observations, they serve as standard rulers for distance measurements, providing a geometric test of cosmological parameters. This approach will provide cosmological information from the shape of the power spectrum transfer function and physically calibrated distance measurements to each redshift shell (e.g., Hu & Haiman 2003, Seo & Eisenstein 2003, Blake & Bridle 2004).

***Supernova luminosity distances:*** In addition to the wide-area survey, the DES will use 10% of its allocated time to discover and measure well-sampled *riz* light curves for ~1900 Type Ia supernovae in the redshift range 0.3<z<0.75 through repeat imaging of a 40 deg$^2$ region. These SNe will provide relative distance estimates to constrain the properties of the dark energy.

In addition to these methods, cross-correlation of CMB data sets with DES galaxies as tracers of potential wells will probe the dark energy through the integrated Sachs-Wolfe (ISW) effect; this effect is included in the forecast constraints below (Hu & Scranton 2004). Finally, we note that accurate photometric redshifts are critical to the DES science goals; as a relatively shallow survey, a major advantage of the DES will be the availability of spectroscopic redshift calibration (training) samples that extend out to the flux limit of the survey.

## 3. Forecast Dark Energy Constraints

In this section, we quantify how the DES will improve our understanding of dark energy, focusing on the dark energy equation of state parameter *w*. Such forecasts generally depend upon priors assumed for marginalized parameters and on assumptions about whether *w* evolves. The marginalized parameters include cosmological parameters other than *w*, uncertain astrophysical parameters that characterize a particular probe, and possible parameters describing uncorrected systematic errors associated with a particular observational method. As a result, caution must be exercised in comparing the projected dark energy sensitivity of different methods and experiments. For this discussion, we assume constant *w* and consider 3 cases of cosmological priors: uniform, present CMB (WMAP 1-year), and future CMB (Planck); these priors are specified in the Supplements. While models with constant $w \neq -1$ are not theoretically well motivated, they nevertheless provide a convenient metric for comparison. A few examples of forecasts with time-varying *w* are discussed in the Supplements. We also note that for varying *w*, for the redshift $z_p$ at which $w(z)$ is best constrained, the constraints on $w(z_p)$ are the same as those on constant *w* shown below.

Table 1: Example forecast marginalized 68% CL statistical DES constraints on constant equation of state parameter *w*.

| Method/Prior | Uniform | WMAP | Planck |
|---|---|---|---|
| Clusters: | | | |
|   abundance | 0.13 | 0.10 | 0.04 |
|   w/ WL mass calibration | 0.09 | 0.08 | 0.02 |
| Weak Lensing: | | | |
|   Shear-shear (S-S) | 0.15 | 0.05 | 0.04 |
|   Galaxy-shear(G-S)+G-G | 0.08 | 0.05 | 0.03 |
|   S-S+G-S+G-G | 0.03 | 0.03 | 0.02 |
|   S-S+bispectrum | 0.07 | 0.03 | 0.03 |
| Galaxy angular clustering | 0.36 | 0.20 | 0.11 |
| Supernovae Ia | 0.34 | 0.15 | 0.04 |

In all cases considered in Table 1, we assume cold dark matter, negligible neutrino masses, adiabatic Gaussian initial perturbations with power-law primordial power spectrum, and a spatially flat Universe. We use a fiducial model with $w = -1$ and other parameters close to the WMAP concordance values. Further assumptions for each method are given in the remainder of this section. The numbers in Table 1 can change as those assumptions are varied within reasonable limits and are meant to be representative. These numbers, based on Fisher matrix and Monte Carlo analyses, indicate that each of the four methods can probe constant $w$ with *statistical* errors at the 3-20% (2-11%) level for WMAP (Planck) priors, assuming reasonable uncertainties in the appropriate astrophysical parameters as noted below. In fact, we expect these methods will likely be limited by systematic errors; a description of their expected impact on the cosmological parameter error budget is presented in the Supplements.

For the cluster results, we have used the cluster counts above the 5σ SPT detection limit (1.9mJy, with a beam of 1' FWHM) in redshift bins of $\Delta z=0.1$ out to $z=1.5$, which results in ~12,000 clusters over 4000 deg$^2$ for the fiducial cosmology and a weakly redshift-dependent mass threshold of ~$2\times10^{14}M_{sun}$. The SZE detection threshold was set this high (as opposed to, say, 3σ) in order to minimize the effects of sample contamination by radio point sources. We have marginalized over a 3-parameter model for the mass-SZE flux relation that includes power-law evolution with redshift beyond that expected from self-similarity, but no scatter in that relation. While this mass-SZE flux relation is rather simple, there is additional information contained in the cluster angular power spectrum and in the shape of the mass function (rather than just its integral above a threshold) that can be used to help ``self-calibrate'' a more complex relation (Majumdar & Mohr 2004, Lima & Hu 2004, 2005). Moreover, the second row of cluster constraints in Table 1 includes the statistical calibration of the mass-observable relation using the weak lensing cluster-shear cross-correlation over the mass range $4\times10^{14}$–$2\times10^{15}$ $M_{sun}$ in redshift bins from $z=0.4$–0.9. Finally, we have assumed that the theoretical uncertainties in the halo mass function, in the halo bias as a function of mass, and in the identification of SZE-detected clusters with dark halos are subdominant compared to the other errors; recent N-body simulations indicate that the first two assumptions are justified and planned future simulations will be needed to ensure the third (see the Supplements for further discussion).

The forecast weak lensing constraints assume that the shear and galaxy power spectra are each measured in 5 photometric redshift bins out to $z=2$ (for background galaxy shear) and $z=1$ (for foreground galaxy positions), with a simplified but reasonable model for the photo-z errors, $\sigma(z)=0.05(1+z)$. The statistical errors come from cosmic variance and from shot noise (shape noise) corresponding to an effective background density of 10 galaxies/arcmin$^2$ (with shape noise per shear component of 0.16); artificially degrading higher resolution images yields this source density for the DES depth and the 0.9'' median seeing delivered by the Mosaic II Camera on the Blanco. If the delivered seeing can be reduced to ~0.7'', close to the median site seeing for the DES observing months, the effective background density will increase by ~35%, with the shot noise errors in the shear-shear power spectrum correspondingly reduced. The results in Table 1 use angular information and assume Gaussian errors up to multipoles $\ell < 1000$, beyond which non-linearities in the density field become important at the typical survey depth; a more conservative (aggressive) limit of $\ell =300$ ($\ell =3000$) increases (decreases) the $w$ constraints by ~50%. In these forecasts, the non-linear mass power spectrum is modeled by the halo model (Hu & Jain 2004), which reproduces the results of high-resolution N-body simulations. For constraints that include foreground galaxies (i.e., G-G and G-S), 5 halo occupation parameters per foreground galaxy photo-z bin are marginalized over. These parameterize the bias of galaxies with respect to the dark matter in a manner consistent with high-resolution N-body simulations (Kravtsov et al 2004) and with the observed clustering of galaxies at low redshift in the SDSS (Zehavi et al 2004). For constraints including (G-G), the foreground galaxy power spectrum is only used to provide constraints on these halo occupation parameters.

The galaxy angular clustering constraint assumes measurement of the angular power spectrum for the foreground galaxy sample with photo-z binning and errors as above. However, it uses a more conservative range of angular information, $\ell < 300$, since baryon wiggles are washed out in the non-linear regime; compared to the first two methods, this result is more robust to uncertainties in non-linear perturbation evolution. Since this clustering constraint mainly uses the shape of the power spectrum, it is not very sensitive to the galaxy bias model. As a result, its use here is complementary to its use above in constraining galaxy bias for lensing.

The forecast supernova constraints assume SNe Ia are standardizable candles with an intrinsic dispersion in peak luminosity of 0.25 mag; this is larger than the usually adopted value of 0.15 mag and reflects an expected increase in errors due to the fact that only photometric redshifts will be available for the majority of the sample. These constraints also assume an irreducible systematic error floor in peak magnitude dispersion of $0.02(1+z)/1.8$ mag in redshift bins of $\Delta z=0.1$ (e.g., Frieman et al 2003). Under this assumption, the error floor dominates over the intrinsic dispersion in the derived dark energy constraints, so there is little gain from reducing the intrinsic dispersion; with no systematic error floor, the $w$ constraints improve to 0.24, 0.14, and 0.03 for uniform, WMAP, and Planck priors. In all cases, we have also assumed a well-measured set of 300 nearby ($z<0.1$) SNe Ia (being accumulated by on-going surveys) anchors the low-redshift part of the Hubble diagram.

**4. Systematic Errors**

Table 2 lists the expected dominant systematic error sources for each method, ordered approximately from most to least important, along with the presently envisioned primary methods for controlling them. A more detailed discussion is presented in the Supplements.

For the cluster technique, the cluster sample must be both complete (above some threshold) and free of contamination, i.e., the cluster selection function must be well understood. For the SZE, cluster selection is complicated by point source confusion, dusty galaxies, radio galaxies, primary CMB anisotropy, and chance projection of clusters at different redshifts. The systematics for DES optical cluster selection are quite different, so the two methods can be compared to understand the selection function. Prior to SPT, the SZA (now operational) will carry out deep SZE imaging over a smaller area of sky with higher angular resolution; this will provide improved calibration of the mass-SZE flux relation and probe the SZE selection function below the SPT threshold. Prior to DES, members of our collaboration will carry out a 100 sq. deg. multi-band imaging survey with Mosaic II on the Blanco (recently approved as a 3-year survey program beginning in Fall 2005) that overlaps several planned SZE surveys (including APEX, ACT, and SPT); this survey will enable cross-comparison of the SZE and optical cluster selection functions for a fraction of the sample. In addition, we will quantify the SZE cluster selection function through a program of hydrodynamic simulations (Melin et al 2004, Vale & White 2005) and Monte Carlo simulations based on radio source catalogs to evaluate contamination. The cluster technique also relies on an accurate mass-observable relation; as noted above, this will be calibrated by statistical weak lensing, the cluster angular power spectrum, and the shape of the cluster mass function.

For weak lensing, the dominant systematic errors are additive and multiplicative shear systematics (Huterer, Takada, Bernstein, & Jain 2005, hereafter HTBJ), uncorrected biases in photo-z estimates (HTBJ 2005; Ma, Hu, & Huterer 2005), and theoretical uncertainties in the small-scale mass power spectrum (White 2004, Zhan & Knox 2004, Huterer & Takada 2004). Theoretical uncertainties can be controlled by nulling the small-scale information (Huterer & White 2005) and by using improved high-resolution N-body simulations incorporating baryons. Photo-z biases will be controlled to an acceptable level by using a pre-existing spectroscopic redshift sample large and deep enough to accurately determine the photo-z error distribution as a function of redshift. The Supplements summarize the acceptable error budget for additive and multiplicative shear systematics (HTBJ) and the techniques we will pursue to reduce them to acceptable levels.

For angular clustering, the dominant systematic errors are potential inadequacy of the halo occupation description of galaxy bias (affecting the baryon wiggles), photometric calibration errors or uncorrected Galactic dust extinction correlated over large scales (leading to artificial large-scale power), and photo-z biases. Since the angular bispectrum and power spectrum have different dependences on the galaxy bias parameters, combining them will constrain those uncertainties (Dolney et al 2004, Gaztanaga & Frieman 1996). In addition, since galaxy bias depends on galaxy type (color and luminosity) measuring angular clustering for different types will constrain the large-scale behavior of the bias. Correlated photometric errors will be controlled by a survey strategy that incorporates multiple visits to each field, by clustering vs. galaxy color, and by checking consistency of results across different angular sub samples.

Table 2: Dominant sources of systematic error and methods for controlling them; see text.

| Method | Dominant Systematic Errors | Primary Controls |
| --- | --- | --- |
| Clusters | Sample selection | SZE + optical cluster selection; simulations |
|  | Mass-observable relation | Self-calibration; statistical WL masses |
| Weak Lensing | Multiplicative shear | Measurement algorithm; shear vs. gal. size |
|  | Additive shear | PCA; active focus; wave-front sensing & alignment control |
|  | Photo-z biases | Spectroscopic calibration sets |
|  | Small-scale power spectrum | Null small-scale power; high-res. simulations |
| Angular clustering | Bias prescription errors | Angular bispectrum; clustering by type |
|  | Large-scale photometric calibration errors | Calibration strategy; clustering by color; angular sub samples |
|  | Photo-z biases | Spectroscopic calibration sets |
| Supernovae Ia | SN evolution | Low and high z SNe comparison |
|  | Photometric errors | Calibration strategy; artificial SNe |
|  | Extinction | SN color and host galaxy information |
|  | Photo-z errors & biases | SN spectroscopic calib. sub sample |

For supernovae, the major systematic errors are evolution of the supernova population, systematic photometric errors, uncorrected host-galaxy extinction, inaccurate K-corrections, and photo-z errors and biases for the part of the sample without spectroscopic redshifts. Evolution is generally controlled by comparing light-curves, colors, and spectra of high- and low-z supernovae and using the fact that the low-z sample, if large enough, should span the range of physical conditions encountered in the sample to z~1. Photometric errors will be minimized by a survey strategy optimized for uniform calibration and cross-checked by overlap with SDSS photometry on the celestial equator. K-correction uncertainties will be mitigated by having a large, nearby comparison SN sample with multi-epoch spectrophotometry. Extinction errors will be addressed by using host galaxy colors to identify a low-extinction sub sample of early-type galaxies. The impact of photo-z errors is discussed in The Supplements and is expected to be small.

**5. Precursor and Concurrent Observations and Developments**

The Dark Energy Survey will make use of a number of other observations and developments that are planned or underway:

1. Spectroscopic redshift data sets to the DES flux limit to calibrate empirical photo-z estimators, to measure photo-z error distributions, and to provide a sample of SN host galaxy redshifts. These will be in place prior to DES from on-going surveys (including SDSS, 2dFGRS, VIMOS VLT Deep Survey, and DEEP2). The overlap of DES with a planned VISTA NIR survey will improve galaxy photo-z estimates but is not required to satisfy the DES science requirements.

2. The SPT survey for SZE measurements of galaxy clusters. SPT and DES plan joint analyses. SPT, which will start survey operations in 2007, expects to have 1-2 years of survey data by the time DES starts operations. A precursor 100 deg$^2$ survey with Mosaic II commencing fall 2005 will overlap several planned SZE surveys, including SPT, and help constrain the cluster selection function.

3. Follow-up spectroscopy of a subsample of ~25% of the SNe Ia on 8m-class telescopes, relying primarily on competitive time applications in collaboration with the supernova community. This will use 8m-class resources at a rate comparable to or less than current high-z SN follow-up; it will reduce cosmological errors from and test the purity of the SN sample. A low-redshift sample of well-measured SNe Ia to anchor the Hubble diagram and provide spectroscopic and photometric templates for SN light-curve fitting and K-corrections will be done by ongoing surveys (KAIT, CSP, SDSS-II, SNF).

4. Suites of large N-body simulations incorporating hydrodynamics by collaboration members to precisely calibrate the theoretical cluster mass function and better model SZE and optical cluster selection. Simulations will also determine with greater precision the effects of clustering non-linearity and baryons on weak lensing and galaxy angular clustering.

## 6. DECam, the Survey Instrument

The philosophy of the DECam project is to assemble proven technologies into a powerful survey instrument and mount the instrument on an optimally configured Blanco, thereby exploiting an excellent, existing facility. Figure 1 shows a cross section of DECam with the key elements identified. A discussion of the Blanco performance and upgrades are given in the Supplements.

The major components of DECam are listed in Section 1 and are summarized below in Table 3. To efficiently obtain *z*-band images for high-redshift (z~1) galaxies, we have selected the fully depleted, high-resistivity, 250 micron thick silicon devices that were designed and developed at the Lawrence Berkeley National Laboratory (LBNL) (Holland et al. 2003). The thickness of the LBNL design has two important implications for DES: fringing is eliminated, and the QE of these devices is > 50% in the *z* band, a factor of ~10 higher than traditional thinned astronomical devices. Several of the LBNL 2k x 4k CCDs of this design have been successfully used on telescopes, including the Mayall 4m at Kitt Peak and the Shane 3m at Lick. The DES CCDs will be packaged and tested at Fermilab, capitalizing on the experience and infrastructure associated with construction of silicon strip detectors for the Fermilab Tevatron program. The CCD packaging plan for the four side buttable 2k x 4k devices builds on techniques developed by LBNL and Lick Observatory.

The optical corrector reference design consists of five fused silica lenses that produce an unvignetted 2.2$^o$ diameter image area, which is calculated to contribute < 0.4" FWHM to the point-spread function. Element 1, the largest, is 1.1m in diameter and the surface of another is aspheric. The spacing between elements 3 and 4 will allow the 600 mm diameter filters to be individually flipped in and out of the optical path. DECam will be installed in a new prime focus cage.

A Fermilab Director's Review (June 2004) and an NOAO Blanco Instrumentation Panel Review (August 2004) evaluated DECam, and both reviews identified the yield of the CCDs, the front end electronics (FEE), and the large optics as the major risks to the project cost and schedule. We have focused our R&D efforts on the mitigation of these risks. The Supplements present further details of the R&D program. In particular, we adopted a proven CCD device design and placed the first DES CCD wafer order. The first test devices were delivered to LBNL in early June 2005 and have been successfully read out on cold probe station. We anticipate delivery of the first thinned fully processed devices this fall. The production of the DES devices by LBNL provides an excellent precursor to the production of devices for the SNAP/JDEM project.

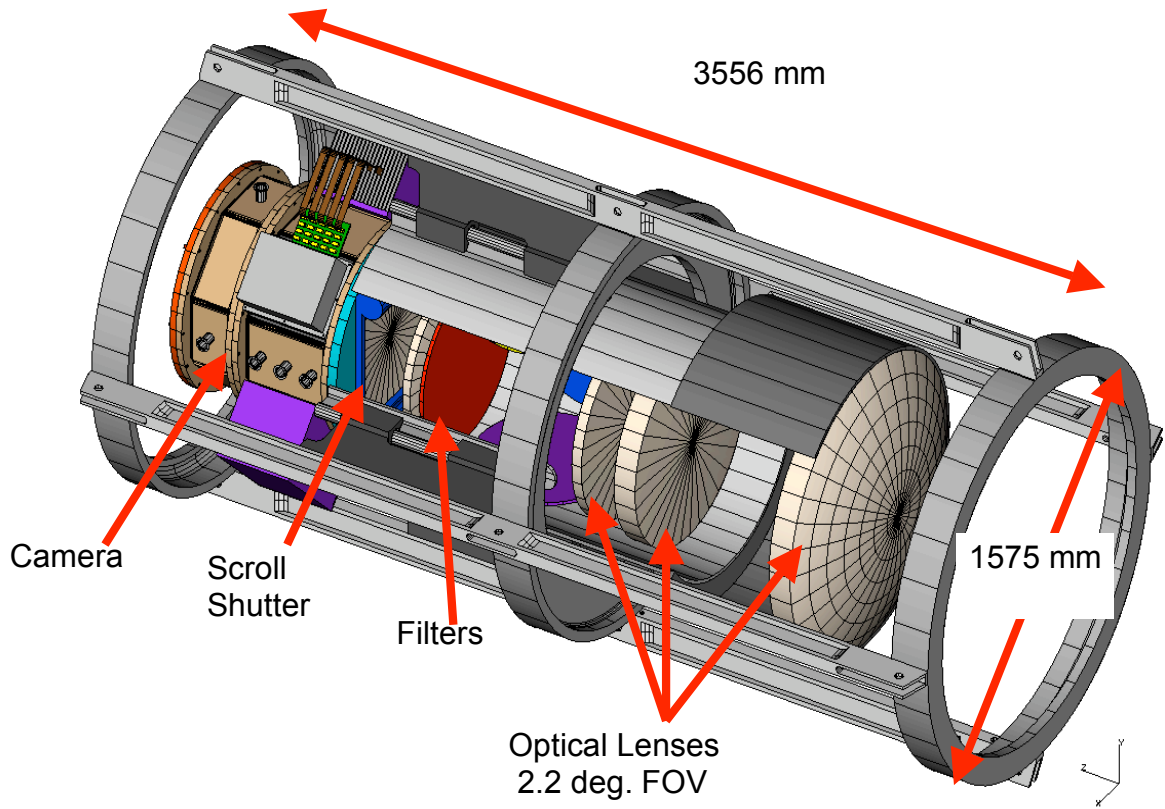

Figure 1: DECam Reference Design

To benefit from the on-going development at NOAO, we have adopted the Monsoon CCD readout system as a starting point. UIUC and Fermilab each have a Monsoon system and are preparing to read out LBNL CCDs in the near future. As we gain experience with Monsoon in the testing setups, we will build on the design and make the modifications necessary to meet the prime focus cage space and heat restrictions.

The risks associated with the optical design result from the size of the elements. We are investigating alternative designs with smaller first elements (~0.9m) and better image quality, with the goal of cost and schedule reduction. We have joined a group organized by George Jacoby to collaborate on the development of large filters for imaging cameras (WIYN, LSST, PanSTARRS). We are also following the development of large colored glass filters at Schott.

Table 3: Expected performance of DECam, Blanco, and CTIO site

| | |
|---|---|
| Blanco Effective Aperture/ f number @ prime focus | 4 m/ 2.7 |
| Blanco Primary Mirror - 80% encircled energy | 0.25 arcsec |
| Optical Corrector Field of View | 2.2 deg. |
| Wavelength Sensitivity | 400-1100 nm |
| Filters | SDSS g, r, i, z |
| Effective Area of CCD Focal Plane | 3.0 sq. deg. |
| Image CCD pixel format/ total # pixels | 2k X 4k/ 519 Mpix |
| Guide, Focus & Wavefront Sensor CCD pixel format | 2k X 2k |
| Pixel Size | 0.27 arcsec/ 15 μm |
| Readout Speed/Noise goal | 250 kpix/sec/ 5 e |
| DECam Corrector (Reference Design) 80% encircled energy (center/edge) | g (0.32/0.59 arcsec) r (0.11/0.37 arcsec) i (0.17/0.41 arcsec) z (0.31/0.47 arcsec) |
| Survey Area | 5,000 sq. deg. |
| Survey Time/Duration | 525/5 (nights/years) |
| Median Site Seeing Sept. – Feb. | 0.65 arcsec |
| Median Delivered Seeing with Mosaic II on the Blanco | 0.9-1.0 arcsec (V band) |
| Limiting Magnitude: 10σ in 1.5" aperture assuming 0.9" seeing, AB system | g=24.6, r=24.1, i=24.3, z=23.9 |
| Limiting Magnitude: 5σ for point sources assuming 0.9" seeing, AB system | g=26.1, r=25.6, i=25.8, z=25.4 |

## 7. Data Management

The DES data management system (DM) is designed to enable efficient, automated grid processing, quality assurance, and long-term archiving of the ~1 Petabyte DES dataset. The raw and processed data will be archived and, after one year, distributed to the public. The survey data will move from CTIO to the National Center for Supercomputing Applications (NCSA) in Illinois, the primary data processing center, over data lines provided by NOAO. The images will be processed, combined into deeper co-added images, and reduced to science-ready data at the catalog level at NCSA. DM is a collaborative effort led by U. Illinois that includes major contributions from Fermilab and the NOAO Data Products Program (DPP). The DM development project will include yearly data challenges that involve testing the system with simulated DES data. Our fourth and final data challenge will end in January 2009, several months before first light for the DES camera.

The DM system includes a pipeline processing environment and data access framework to enable automated and modular processing of this large dataset. This framework will be provided by NCSA and is closely coupled to their large, middleware development effort for the LSST data management project. The DM system includes astronomy modules for processing and data quality assurance, which will come from the collaboration. The primary image archive will employ the NOAO Science Archive software, which is being developed by NOAO DPP. The development of the DES catalog database and server is being led by NCSA.

## 8. DES Timeline

| | |
|---|---|
| October 2004 | Start DECam R&D and continue the preliminary design |
| April 2006 | Hold preliminary design review, obtain DOE project approval, and place long lead procurements with non-DOE funds |
| October 2006 | Place long lead procurements with DOE funds, begin production processing, packaging and testing of CCDs |
| October 2008 | Complete construction of DECam and Data Management System (DM) |
| February 2009 | Deliver DECam and DM to CTIO |
| May 2009 | Begin commissioning of DECam on the Blanco with the completed DM |
| September 2009 | Begin observations |
| March 2014 | Complete observations |

## 9. Conclusions

The Dark Energy Survey will employ four complementary techniques to study dark energy: galaxy clusters, weak lensing, galaxy angular clustering, and supernova distances. The statistical reach of these techniques is well understood; in the DES, *each* of them will deliver statistical constraints on dark energy that are stronger than the best *combined* constraints available today (Spergel et al 2003, Tegmark et al 2004, Seljak et al 2004). Moreover, our collaboration is making substantial progress toward identifying and understanding the dominant astrophysical uncertainties and observational systematic errors for each of these methods and one of our important goals is to further explore and develop methods to control these systematic errors. Since the more ambitious surveys of the future will reach even smaller statistical errors than the DES, they will have to exercise even finer control of systematic errors in order to achieve their science goals. We believe that a large-scale, near-term survey that provides a major step forward in precision such as DES is the logical next step in that process.

The DES will employ DECam, a powerful new wide-field survey instrument, and the Blanco, a 4m telescope that has already contributed to many of the pioneering measurements of dark energy and that has the capacity for improvements that will strengthen the DES measurements. As a relatively shallow survey, the DES makes use of source galaxies that are large enough to be well resolved in the conditions routinely achieved with MOSAIC II, the current Blanco imager, and bright enough so that their photometric redshifts can be well calibrated by spectroscopic surveys of comparable depth. The collaboration institutions have a proven record in astronomical data management and have the capacity to manage large data sets. Collaboration members have made important contributions to developing the survey science, and include a strong science team that will rise to the challenge of carrying out the astrophysical and cosmological simulations that will be needed to precisely interpret the data from this large survey. The DES promises significant scientific returns, although it is a relatively low-risk project of intermediate scope and cost, which requires only modest advances beyond the hardware and software used in current astronomical projects.

DES will also provide the astronomical community with a wide field, 4 band digital survey of the southern sky with excellent image quality, uniform photometry and unprecedented depth for its sky coverage. It will cover the largest volume of the universe to date (complete to tens of $Gpc^3$) and it will be a "legacy survey" that will provide the scientific and educational communities with an extraordinary catalog for multipurpose projects.

The DES and the SPT projects provide a unique opportunity to combine two strong surveys into a single survey that will be greater than the sum of its parts. The very strong impact that they can make together on cosmology will be much greater if the observations are made in a timely way. The SPT project will begin observations in 2007, thus it is important for DES to start its build phase soon.

# The Supplements for the Dark Energy Survey White Paper

These Supplements provide additional details, background, and context for the DES White Paper. The first section describes the cosmological parameter priors used in the forecasts of Table 1 in the White Paper. The next four sections describe the four primary techniques that we plan to use to measure dark energy: galaxy clusters, weak lensing tomography, galaxy angular clustering, and supernovae distances. Section 6 describes plans to obtain photometric redshifts and to understand the photo-z error distributions, and Section 7 describes the program of simulations. Section 8 describes the improvement program for the Blanco telescope that we plan to carry out. Section 9 provides more information on the R&D program that we have initiated for DECam.

## 1. Priors on Cosmological Parameters

The forecast dark energy constraints in Table 1 in the White Paper were carried out for 3 different assumptions about prior information on the remaining cosmological parameters. The analysis adopted here was restricted to a simple set of cosmological parameters: in addition to the dark energy equation of state $w$ (assumed constant), they are the physical baryon density ($\Omega_b h^2$), the physical matter density ($\Delta\Omega_m h^2$), the angular size of the last scattering surface ($\theta_{LSS}$) is a more suitable parameter for CMB measurements than the Hubble parameter, the optical depth $\tau$ due to reionization, the amplitude of scalar density perturbations ($A_s$), and the power-law index of the primordial power spectrum ($n_s$). We assume Gaussian, adiabatic initial perturbations with constant (non-running) $n_s$, cold dark matter, massless neutrinos, no tensor modes, and a spatially flat Universe. For the 3 cases, the priors on these 6 parameters were chosen as follows:

I. A uniform prior on the relevant cosmological parameters.
II. WMAP priors: The weak lensing, galaxy angular clustering, and supernova constraints were obtained from the Fisher projection for one year of WMAP data using the noise from the LAMBDA site (http://lambda.gsfc.nasa.gov/). The covariance matrix from an analysis of the WMAP data with a fixed $w$ was used for the cluster number counts in this case, the marginalized errors on the parameters are: $\Delta\Omega_b h^2$ = 0.0025, $\Delta\Omega_m h^2$ = 0.017, $\Delta\theta_{LSS}$ = 0.0081, $\Delta\tau$ = 0.12, $\Delta n_s$ = 0.07, $\Delta[\log(10^{10} A_s)]$ = 0.24.

III. Planck priors: The Fisher projection of all Planck LFI and HFI channels for weak lensing, galaxy angular clustering, and supernovae,. For cluster number counts, the Planck prior is given by a full covariance matrix (Lewis 2005); we did not include the effects of CMB lensing, and a fixed $w$ was assumed. The effective observation was taken to be 1.5 years of data with an effective 7' beam up to $\ell_{max}$ = 2000. In this case, the marginalized errors on the parameters are: $\Delta\Omega_b h^2$ = 0.00014, $\Delta\Omega_m h^2$ = 0.0014, $\Delta\theta_{LSS}$ = 0.00029, $\Delta\tau$ = 0.0053, $\Delta n_s$ = 0.0037, $\Delta[\log(10^{10} A_s)]$ = 0.010.

In all cases, we choose a fiducial model with $w = -1$ and other parameters set at or near the WMAP concordance values. We note that the CMB-only constraints on the equation of state are quite weak, $\Delta w$=0.8 (WMAP) and 0.3 (Planck); as Fig. 5.1 shows, the CMB constrains a degenerate combination of $w$ and $\Omega_m$.

## 2. Galaxy Clusters

The formation and evolution of structure is seeded by initial perturbations and driven by gravitational instability in a dynamically evolving spacetime. Massive structures observed in the Universe today bear the marks of these three influences: initial perturbations, the processes of gravitational collapse, and the evolving underlying metric. Galaxy clusters, the largest virialized objects in the mass distribution, are a particularly tractable target for observations of structure and its evolution over cosmic time.

*Method:* The observed cluster redshift distribution is the product of the comoving volume per unit redshift and solid angle, $d^2V/dzd\Omega$, and the comoving density of detected clusters $n_{com}$,

$$\frac{d^2 N}{dzd\Omega}(z) = \frac{d^2 V}{dzd\Omega}(z) n_{com}(z) = \frac{c}{H(z)} D_A^2 (1+z)^2 \int_0^\infty dM\, f(M,z) \frac{dn}{dM}(z)$$

where $dn/dM$ is the cluster mass function, $H(z)$ is the Hubble parameter as a function of redshift, $D_A(z)$ is the angular diameter distance, and $f(M,z)$ is the redshift-dependent mass selection function of the survey which is zero below some limiting mass. The cosmological sensitivity comes from three basic elements:

- **Volume**: the volume per unit solid angle and redshift depends sensitively on cosmological parameters.
- **Abundance Evolution**: the evolution of the number density of clusters, $\frac{dn}{dM}(z)$, depends sensitively on the growth rate of density perturbations, which is determined by the expansion rate $H(z)$ and therefore highly sensitive to cosmological parameters.
- **Mass selection function**: clusters are selected using an observable correlated with cluster mass. In general, the observables are used in a way that is cosmology dependent. For the SPT, the observable is the SZE flux decrement.

The cluster counting method depends critically on accurate statistical cluster mass estimates and on well-characterized completeness of the cluster samples as a function of redshift. Observations to date indicate that the SZE flux is strongly and robustly correlated with total cluster mass. The DES will provide statistical cross-calibration of SZE cluster mass estimates using the cluster-mass correlation function inferred from weak lensing. In addition to measuring the photometric redshifts of SPT clusters, the DES will detect clusters optically to lower mass thresholds than the SPT using the red galaxy color-magnitude sequence (Gladders and Yee 2000); this will provide a cross-check on the SZE cluster selection function as well as an important independently selected cluster sample in its own right. Moreover the DES will trace the evolution of the stellar-phase baryons in the cluster and provide a further mass estimator. By coordinating the DES with the SPT survey, we will cross check cluster selection and mass estimates and thereby better control systematic errors.

***Constraints:*** In our calculations of the constraints on dark energy from SPT+DES, we assume the SPT survey has a $\theta_{FWHM} = 1'$ beam, one channel at 150 GHz, and a 1.9mJy ($5\sigma$) detection threshold over 4000 deg$^2$. We include the effects of clusters larger than 1' by degrading the beam size, up to a maximum of 5' (Battye and Weller 2003). Using a standard $\beta=2/3$ gas density profile with a central overdensity factor of 10 and a maximum redshift $z_{max} = 1.5$ results in ~12,000 SPT clusters. The predicted SZE cluster mass limit and cluster abundance as a function of redshift is shown in Figure 2.1 for the fiducial cosmology.

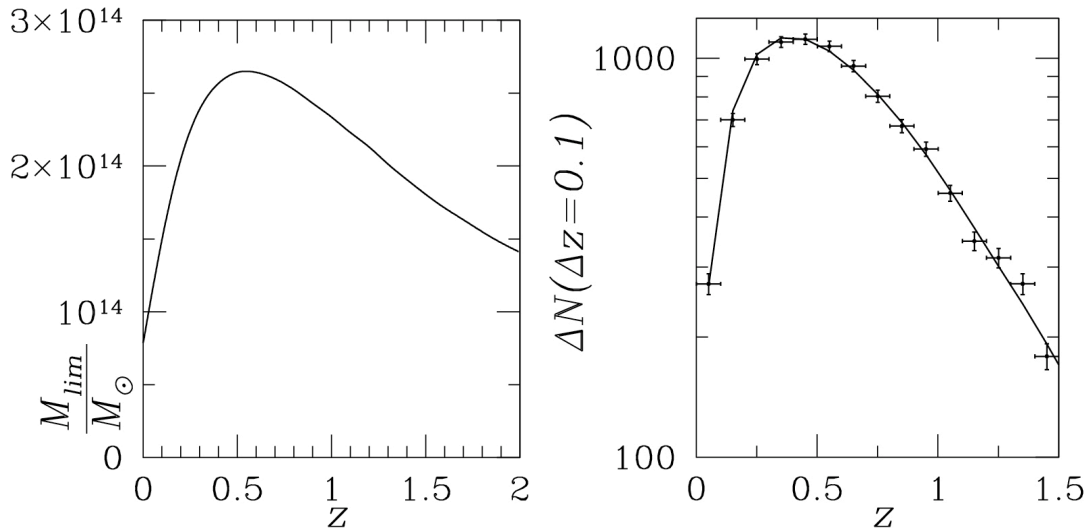

Figure 2.1. The predicted SZE $5\sigma$ cluster mass limit and abundance above that limit as a function of redshift for the SPT+DES survey for the fiducial cosmology used in the White Paper.

In this analysis, the effects of point sources, both infrared galaxies and radio AGN, are not included, although White and Majumdar (2004) have shown how to do this for a single frequency instrument. SPT will have three frequency channels and can address point source confusion by applying the spectra of the point sources and hence cleaning the map. Here we have addressed this issue qualitatively by using a $5\sigma$ (rather than lower) detection threshold for SPT.

***Systematic Errors:*** As Table 1 of the White Paper indicates, the SPT+DES cluster survey can achieve high statistical precision in constraining dark energy. To approach that precision, we need to understand the sources and expected levels of systematic error and develop methods to control them to the extent possible. We will present an inventory of the systematic errors by stepping through the program, starting with the theoretical predictions and continuing to the observations.

A key strength of the DES is the ability to apply multiple techniques for both identifying clusters and inferring their masses, each with their own systematics. The SZE method using SPT (our primary method) has a theoretically clean mass-observable scaling relation, in which the observable is integrated SZE flux, and precursor surveys such as the SZA will help improve its accuracy. Much of the remaining work will be focused on quantifying systematics in the cluster survey selection function. The optical cluster method has a relatively clean selection function, at least for clusters more massive than $10^{14} M_{sun}$, but work remains to be done on quantifying systematics in the optical mass-observable scaling relations. Shear-selected cluster samples, which DES will also provide, are in principle statistically powerful probes, but there are substantial challenges to accurately modeling the selection function, primarily due to projection effects associated with the broad redshift range of the lensing kernel and the intrinsic noise that prevents identification of low-mass objects. We will pursue all three of these techniques, and the cross-checks they provide will be an important check on the different methods.

*Theory:* Given an underlying cosmology and a spectrum of initial perturbations, the evolution of the massive peaks in the density field can be calculated, either analytically (as in Sheth and Torman, 2002) or numerically (Jenkins et al 2001; Evrard et al 2002). Predictions for the abundance and clustering of massive halos are becoming increasingly accurate; different methods currently agree to better than 10%. Work in the community on N-body simulations incorporating dark energy, higher mass resolution, and hydrodynamics continues. The current state of the art is the careful study of the mass function by Warren et al. (2005). In the future, we plan to carry out large suites of high-resolution simulations that include gas dynamics in order to precisely quantify the theoretical predictions, including the theoretical uncertainties, and to improve the understanding of cluster selection in the different methods. Based on the results to date, it appears that the theoretical prediction of the cluster mass function to sufficient accuracy for the needs of our survey is a manageable problem.

*Scaling relations:* The cluster mass function is not directly observable, and the predictions must be made in terms of observable quantities. Fortunately, clusters are richly labeled objects and many of their observational properties correlate well with mass. With appropriate observations one can measure X-ray properties ($L_x$ and $T_x$), galaxy and stellar populations ($N_{gal}$, $L_{tot}$, etc.), and Sunyaev-Zeldovich distortions of the microwave background. Understanding the relationship between these observables and the appropriate mass that enters the mass function is central to cluster cosmology. We need to measure the mass scaling relations and their scatter.

Our primary technique for handling the mass scaling relations takes advantage of the various ways that cluster samples provide for cross checking mass estimates (e.g., Majumdar and Mohr 2004). Given photometric redshift errors, there are ~5 useful redshift shells in the DES; in each redshift bin, the mass function shape and amplitude and the spatial correlation function must be predicted by a successful cosmology and scaling relations, including scatter and evolution. Not all of these pieces of information need be used, and in fact it is often useful to hold one back as a final cross check. This cross checking of the mean mass estimate is required because no technique produces an absolutely clean low mass threshold: depending on the scatter in the scaling relation, clusters with mass below the putative mass threshold are included in the sample and a smaller number with mass above the threshold are scattered out. Theoretically this is a source of uncertainty in the predicted numbers of clusters; in practice we find that the cluster samples themselves provide enough observables to constrain the parameters that cause the threshold to be fuzzy, i.e., to self-calibrate the mass function.

*Scaling relations: SZ-* Current simulations suggest a small scatter of order 10% in the SZE flux vs. mass relation within $r_{200}$ and $M_{200}$ (de Silva et al 2004, Motl et al 2005). Observational samples are still small, but investigations have begun, for example Benson et al (2005), LaRoque et al (2005). The SZA cluster survey, coupled with high-resolution simulations, will enable us to measure the shape, evolution, and scatter in this relation with higher precision.

*Scaling relations: Optical-* Investigations using simulations to address the scaling relations between, e.g., mass and richness ($N_{gal}$) or mass and optical luminosity ($L_{opt}$) have begun (Weschler et al 2005; Miller et al 2005), and suggest a well defined relationship with a mass dependent scatter of about 40% at masses > $1\times10^{14}$ solar mass. The observations of massive clusters indicate a well defined scaling relation (Gilbank et al 2004; Lin, Mohr, & Sanford 2004; Yee & Ellingson 2004); Popesso et al (2005) report a scatter of 20% for $L_{opt}$-$M_{500}$. We will carry out high-resolution simulations and develop halo occupation distribution function modeling to make further progress in understanding the dispersion in mass as a function of the observables.

As it is possible to bin the cluster sample on the observable $N_{gals}$ or $L_{opt}$, it is possible to reduce another systematic error. The relevant mass is that which enters into the mass function. The mass function predictions are most often in terms of the virial mass or the mass inside $r_{200}$ (the radius inside which the mass density is 200 times the critical density). This radius can be found statistically for the observed cluster samples by stacking the sample in bins of $N_{gals}$ or $L_{opt}$ and studying the run of the observable with radius. Hansen et al (2004) used an observable overdensity, the radius inside which the galaxy density is 200 times the background, while Johnston et al (2005) have developed a method to invert the mean 2-d weak lensing shear profile into a 3-d mass profile, from which $r_{200}$ and virial mass can be read off. Both of these observational determinations can be calibrated against appropriate simulations.

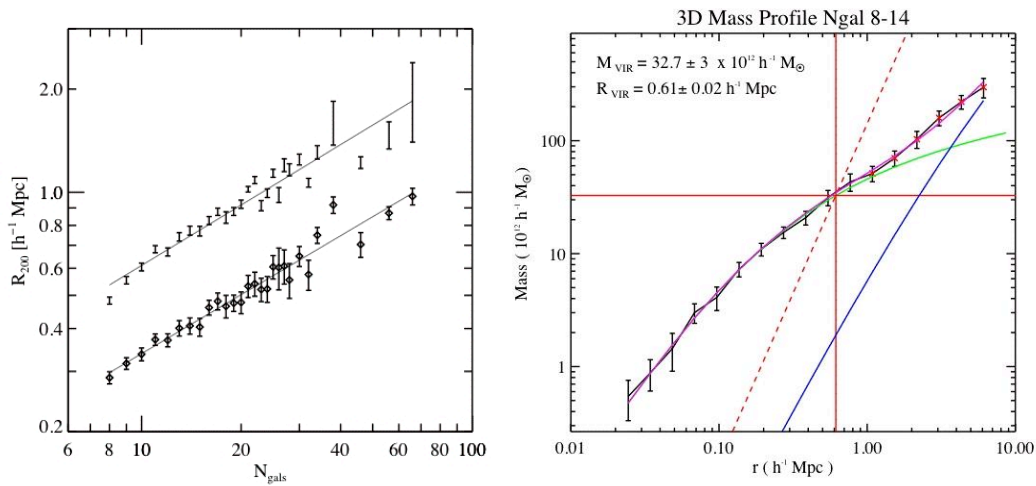

Figure 2.2. Relationship of the cluster characteristic radius $r_{200}$ to richness. *Left panel*: mean radius at which the galaxy density is 200 times the critical (diamonds) or mean (bars) density, as a function of optical cluster richness, $N_{gals}$ (Hansen et al. 2004). *Right panel:* mean cluster mass profile deprojected from cluster-shear measurements for clusters identified optically and containing between 8 and 14 galaxies brighter than $L_*/2$ along the red cluster sequence (Johnston et al 2005); vertical red line denotes the virial radius and green curve shows the best-fit NFW profile for the points inside the virial radius. Both of these studies use SDSS data and are at $z\sim0.2$.

*Scaling relations: Cross Check* - The DES data will allow weak lensing measurements of the clusters, either in the form of the cluster-shear correlation function or individual noisy mass estimates. This enables the DES to provide an independent cross-check on the SZE cluster mass estimates statistically. As the cluster counting method depends critically on the accurate statistical mass estimation, we emphasize how powerful it is to have this independent cross check.

*Selection Functions: SZE*- The systematic effects in the selection of cluster samples from the SZE survey are primarily from sources which contaminate the SZE flux measurements. Much work is currently on-going to understand these issues in greater detail (e.g., Schultz and White, 2003; White and Majumdar, 2004; Melin, Bartlett, and Delabrouille 2005). The two most uncertain effects are radio loud AGN, potentially associated with the clusters themselves, and sub-mm bright galaxies, high-redshift galaxies either undergoing bursts of star formation or harboring an AGN but in either case enshrouded by dust (White and Majumdar 2004). Multi-frequency observations such as those planned for SPT can alleviate this problem. These contaminants will be better understood in part by pursuing a program of optical identification of clusters in small-area, high-resolution SZE surveys that will take place in the next few years, e.g., with the SZA.

*Selection Functions: Optical-* The selection function in the optical is quite benign, as the clusters of primary interest for cosmology, at masses of $> 2\times10^{14}$ $M_{sun}$ have >30 luminous red galaxies within 1 Mpc of the cluster center. These clusters are thus easy to detect. The primary systematic in the optical catalogs is the need to understand the relationship between luminous galaxies and mass at all redshifts. This relationship varies with redshift due to both intrinsic evolution of the galaxies and observational shifts due to the K-correction. Since clusters are dominated by a population of early-type galaxies with very uniform, old stellar populations, this problem is relatively tractable. The evolution of these galaxies through the redshift range of interest is increasingly well explored observationally, and the galaxy evolution models becoming ever more constrained. Members of the DES collaboration are pursing this investigation using both SDSS and RCS data sets. Of lesser concern is the inability to separate clusters that are projected on the sky at redshift separations of $\delta z < 0.03$, as this can be corrected for given the simulations; in any event, this is relatively benign, since the effect of producing apparently more massive systems through projection is balanced by reducing the total counts as two or more systems are treated as one.

*Selection Functions: Cross Check-* The cluster counting method also depends critically on well-characterized completeness of the cluster samples as a function of redshift. The DES will detect with high completeness the optical clusters to a much lower mass threshold than the SZE clusters of the SPT survey, and with a selection function with very different systematics. Once again we emphasize how powerful it is to have cross checks on the primary areas where systematic errors are thought to be an issue.

## 3. Weak Gravitational Lensing

The bending of light by foreground mass concentrations shears the images of distant source galaxies. Dense mass concentrations such as galaxy clusters induce a coherent tangential shear pattern that can be used to reconstruct their surface mass densities. Larger scale structures with lower density contrast also generate correlated shear, but with lower amplitude—in this case one studies the shear pattern statistically by measuring shear-shear correlations (also known as cosmic shear). Since the foreground dark matter is associated to large degree with foreground galaxies, one can also measure the angular correlation between foreground galaxy positions and source galaxy shear (galaxy-shear correlations, also called galaxy-galaxy lensing). These weak lensing techniques provide powerful probes of the dark energy in the context of the Dark Energy Survey. In addition, as noted in the previous section, lensing will provide statistical cluster mass estimates through the cluster-shear correlation function that can cross-check SZE and optical richness mass estimators.

The shear-shear, galaxy-shear, and galaxy angular power spectra can be expressed as projections of the corresponding three-dimensional power spectra (e.g., Hu & Jain 2003),

$$C_\ell^{x_a x_b} = \int dz \frac{H(z)}{D_A^2(z)} W_a(z) W_b(z) P^{s_a s_b}(k = \ell/D_A; z)$$

where $\ell$ denotes the angular multipole, $a, b = (1, 2)$, $x_1$ and $x_2$ denote the two-dimensional angular galaxy (*g*) and shear (*γ*) fields, and $s_1$ and $s_2$ respectively denote the three-dimensional galaxy (*g*) and mass (*m*) density fluctuation fields at redshift *z*. The weight functions $W_1$ and $W_2$ encode information about the galaxy redshift distribution and about the efficiency with which foreground masses shear background galaxies as a function of their respective distances.

The dark energy density and equation of state affect these angular power spectra through the Hubble parameter, the angular diameter distance, the weight factors, and through the redshift- and scale-dependence of the three-dimensional power spectra $P^{gg}$, $P^{mm}$, and $P^{gm}$. For a given choice of cosmological parameters, the mass power spectrum $P^{mm}$ can be accurately predicted from N-body cosmological simulations down to some mass scale; the shape (scale-dependence) of $P^{mm}$ is also well constrained on large scales by CMB anisotropy data. In addition to cosmology, the power spectra involving galaxies, $P^{gg}$ and $P^{gm}$, require a model for the *bias*, that is, for how luminous galaxies are distributed with respect to the dark matter. Since galaxy bias is a complex process, we have based our forecasts on a conservative analysis that marginalizes over 5 free parameters related to bias in each redshift bin (25 bias parameters in all). These parameters are based on the 'halo occupation distribution' model (see Hu & Jain 2003 for details) that describes how galaxies above some luminosity occupy dark matter halos as a function of halo mass. For the forecasts in Table 1 of this White Paper, we use a minimum halo mass of $10^{12.5}$ $M_{sun}$. Constraints on these bias parameters come mainly from the galaxy angular power spectrum measurement, since it has the highest signal to noise of the three. This bias model is physically motivated, accurately reproduces the results of N-body simulations that include gas dynamics or that resolve sub-halos, and accounts for the galaxy two-point correlation function measured as a function of luminosity in the SDSS (Zehavi, etal 2004).

To forecast constraints, we estimate the statistical errors on the angular power spectra; for example, for the shear-shear spectrum, the uncertainty is (Kaiser 1992)

$$\Delta C_\ell^{\gamma\gamma} = \sqrt{\frac{2}{(2\ell+1)f_{sky}}} \left( C_\ell^{\gamma\gamma} + \frac{\sigma^2(\gamma_i)}{n} \right)$$

where $f_{sky}$ is the fraction of sky area covered by the survey, $\sigma^2(\gamma_i)$ is the variance in a single component of the (two-component) shear, and $n$ is the source galaxy angular number density per sr. The first term in brackets comes from cosmic variance, and the second, shot-noise term results from both the variance in galaxy ellipticities ('shape noise') and random error in measuring galaxy shapes. This expression assumes the shear field is Gaussian; although this assumption breaks down at large $\ell$ due to non-linearity, the non-Gaussian variance is generally masked by the shape noise term. In the Table 1 forecasts, we only use information from multipoles $\ell$<1000, so the maximum spatial wavenumber used is $k\sim 1$ $h^{-1}$ Mpc at the typical depth of the survey.

The variance expression indicates that weak lensing places a premium on maximizing the survey sky coverage and the surface density of source galaxies with measurable shapes. We rewrite the shot-noise term as $(0.16)^2/n_{eff}$, where the numerator is the empirical shape noise for large, well-measured galaxy images (Sheldon et al 2004, Jarvis et al 2003), and $n_{eff}$ is the source galaxy density including noise weighting. In estimating the shear, the ellipticity measurement of each source galaxy is relatively weighted by the inverse noise, which has contributions from shape noise and shape measurement error. In addition, the ellipticity of each galaxy is corrected for PSF dilution by a factor that depends on the square of the ratio of the galaxy size to the PSF—for small galaxies the correction is large, and uncertainty in the correction factor means these galaxies are further down weighted in the shear estimate. These effects are incorporated into the noise-weighted galaxy density $n_{eff}$. Based on scaling from CFHT12K data (in excellent seeing) and HST GOODs data, we estimate $n_{eff} \approx 10$ arcmin$^{-2}$ for the completed DES survey (integrated exposure times of 400 s in *r,* 1200 s in *I,* 2000 s in *z*), assuming median delivered seeing of 0.9". This seeing value is consistent with the median delivered to the Mosaic II for the observing months of interest. In fact, the delivered seeing for DES should be better than this, due to a new optical corrector designed to deliver smaller PSF, active control of the focus, and anticipated improvements in the thermal environment of the prime focus cage and dome. A decrease in seeing would have two effects: (i) the PSF is a smaller fraction of the size of the fainter source galaxies, so they receive more weight due to the smaller PSF dilution factor; (ii) the effective number of CCD pixels per object is smaller, decreasing the sky background per object and therefore the shape measurement error. We find that $n_{eff}$ increases by approximately 17% per 0.1" reduction in seeing around our fiducial seeing value. Finally, we note that the galaxy-shear correlation function has been measured with high signal to noise in the SDSS (Sheldon, etal 2004), with typical seeing of 1.5" and source density of only $n_{eff} \sim 2$ per square arcmin. By comparison, the DES will cover a similar area on the sky but go considerably deeper with substantially better seeing.

The last row in the weak lensing section of Table 1 of the White Paper gives the improved constraints on *w* if information from the cosmic shear bispectrum is included along with shear-shear correlations. For this forecast, we have combined information from triangles of all configurations and tomographic bins (Takada & Jain 2004). By the time of the survey, we intend to verify the analytical methods used here with ray tracing simulations. In particular we will include various higher order effects (Dodelson & Zhang 2005) and the non-Gaussian covariance between the power spectrum and bispectrum; initial results indicate that the impact on parameter constraints will not be significant (e.g. Kilbinger & Schneider 2005). We also note that the bispectrum can alternatively be used as a check on systematics in the shear power spectrum.

It is interesting to consider how the WL constraints in Table 1 change as we vary priors and model assumptions. For example, we can drop the prior from the CMB constraint on the angular diameter distance to the last-scattering surface, which constrains a degenerate combination of *w* and $\Omega_m$. This is roughly equivalent to dropping the assumption of spatial flatness; in this case the shear-shear and galaxy-shear+galaxy-galaxy errors on *w* in Table 1 increase by less than ~40% for either WMAP or Planck priors, showing that these constraints are robust to spatial geometry. Finally, we can drop the assumption of constant *w* and consider constraints on the time variation of the equation of state, *dw/da*. The resulting shear-shear constraints on *dw/da* are 1.3 (uniform prior), 0.5 (WMAP), and 0.4 (Planck); for galaxy-shear+galaxy-galaxy, the corresponding constraints are 0.8 (uniform), 0.4 (WMAP), and 0.3 (Planck).

*Weak Lensing Systematic Errors:* The statistical accuracy of the lensing constraints in Table 1 can only be approached if systematic errors are kept under control. Systematic errors in weak lensing measurements can arise from incorrect shape measurements and biases in the photometric redshifts of galaxies. Shape measurement errors result from incorrect calibration or errors in correction of the anisotropic point spread function (PSF) and lead to erroneous multiplicative or additive factors in the estimated lensing shear, respectively. The lensing shear we wish to measure is ~1% on arcminute scales and drops to ~0.1% on degree scales; the goal is to reduce the sum of all systematic errors below statistical errors over at least this range of angular scales.

Although the shear-shear spectrum is proportional to $P^{mm}$ and thus independent of the halo occupation parameters, it is the most sensitive to systematic errors in measuring galaxy shapes. The galaxy-shear correlations are less sensitive to these systematics than shear-shear correlations, because PSF anisotropy tends to cancel out of the azimuthally averaged tangential shear field measured around foreground galaxies. The foreground galaxy angular power spectrum is independent of the shear systematics but is more sensitive to uncertainties in the halo occupation prescription for bias (see section 4).

Given the subtle nature of lensing systematics and the novelty of analysis techniques in this field, the DES is a low-risk survey because it is relatively shallow and the Blanco telescope is well studied. The shallow depth of the survey means the lensing signal will be lower, but more important, the relatively large angular size (compared to the seeing disk) and higher surface brightness of the source galaxy population will simplify considerably the correction of systematic errors. Given the typical delivered seeing at the Blanco, the depth versus area trade-off of DES is well suited for obtaining the best possible lensing measurements. The 75 square degree lensing survey of Jarvis et al (2005) using the Blanco telescope has comparable depth to DES in the *r*-band; their results demonstrate the feasibility of the planned measurements with DES.

Finally, for the expected redshift distribution for DES, photometric redshifts can be measured robustly and calibrated with spectroscopic samples that reach the same flux limits--this is important for interpreting the lensing measurements.

*Shape measurement error:* The dominant shape measurement error in current lensing surveys is due to the anisotropy of the point spread function (PSF), caused by optical and CCD distortions, tracking errors, wind shake, atmospheric refraction, etc. In a given exposure, the PSF anisotropy is measured using the stars in the field and interpolated to the positions of the galaxies. Because the density of stars is relatively low, the errors in this interpolation can lead to significant coherent errors in the measured shapes of galaxies that are difficult to distinguish from the lensing shear. This leads to additive errors in the shear that must either be eliminated or marginalized over. Since the PSF anisotropy depends on the detailed optical-mechanical properties of a telescope as well as observing conditions, it is specific to each telescope and observing site. The error in measuring the PSF anisotropy can be reduced by combining data from multiple pointings of a telescope, increasing the effective density of stars. However, time variability of the PSF shapes limits the effectiveness of this technique. Recently the error in measuring the PSF shape has been substantially reduced by introducing a Principal Component Analysis (PCA) technique that optimally uses information on the PSF from multiple exposures (Jarvis & Jain 2004); this enables interpolation of the PSF with much finer effective angular resolution. This is especially promising for the DES, since it has already been applied to data taken with the BTC and Mosaic II imagers on the Blanco telescope.

We have examined the principal components of the distortion patterns of data from the Blanco and made forecasts for how well the DES can correct these. From the current design for the DES optical corrector, we expect the level of raw PSF anisotropy to be a few percent on arcminute scales and have estimated that with the PCA technique we will be able to reduce the measurement error by over a factor of 100. This is based on the analysis of Jarvis & Jain (2004), scaled conservatively to the number of exposures for DES (the correction improves with the number of exposures). In this extrapolation we have accounted for the number of well-measured stars available to estimate the principal components. The estimated residuals are well below the statistical errors. Hence we believe that additive errors in the estimated shear will be negligible in the error budget of DES lensing measurements.

The second kind of shape measurement error arises due to calibration errors and contributes a multiplicative term to the shear; it can arise, for example, from inaccurate correction for the circular blurring of galaxy images (and consequent reduction in ellipticity) due to seeing. The finite size of the PSF and the distribution of intrinsic shapes of galaxies need to be accurately measured to calibrate the shear. This has been done in recent studies by a careful combination of analytical techniques and tests with simulated images. Initial results from an ongoing set of tests known as the STEP program (Heymans et al 2005) suggest that software based on the Bernstein & Jarvis (2004) approach can achieve a calibration accuracy approaching the requirements for DES. More work is needed, and the results must be validated with more extensive tests, which we intend to pursue.

It is further reassuring that calibration errors appear to be the least dangerous systematic for cosmological measurements (Huterer et al 2005). Unlike additive errors, with calibration errors there is less freedom in mimicking the redshift dependence of the signal. Hence even if the calibration errors increase from 0.4% to 4% of the shear, Fig. 3.1 shows that the resulting constraint on $w$ degrades by less than a factor of two. Moreover, the shear measurements can be studied as a function of source galaxy angular size; since the calibration uncertainties go down as the square of the PSF divided by galaxy size, the consistency of the shear signal as a function of source galaxy size provides a check on these calibration errors.

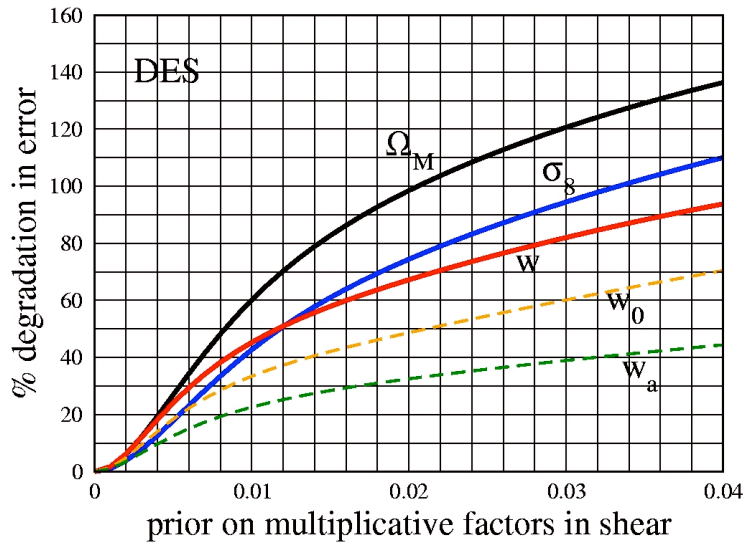

Figure 3.1. Degradation in error on cosmological parameters due to marginalizing over uncorrected multiplicative shear systematic errors, from Huterer et al. (2005).

*Photometric redshift errors:* The impact of systematic photometric redshift errors on shear-shear measurements in the DES have been studied by Huterer, et al. (2005) and by Ma, Hu, & Huterer (2005); for more discussion, see Section 6 below. For source galaxies binned by photometric redshift, they find that the error (bias) in the centroids of the photo-z bins must be measured at the level of 0.002 or less and the scatter in the photo-z errors per bin known to an accuracy better than 0.01 in order to keep the lensing constraint on $w$ from degrading by more than 10%. Using a subsample of galaxies with measured spectroscopic redshifts, the photo-z error distribution as a function of redshift can be measured, and this scatter and bias inferred, with an uncertainty depending on the size of the spectroscopic sample. For the characteristic photo-z errors predicted by the DES photo-z simulations described in Section 6, Ma, Hu, & Huterer (2005) show that a spectroscopic `training set' that reaches the DES photometric depth and comprises 50,000-100,000 galaxy redshifts will achieve these desired accuracies. As noted in Section 6, on-going deep spectroscopic surveys will provide such a sample before DES begins taking data, and we therefore expect photo-z systematic errors to be under control for DES weak lensing.

***Optical performance of the Blanco telescope:*** In order to better understand the telescope optical and mechanical properties and how they may impact the DES, we have begun a detailed analysis of the optical distortions of the Blanco using ray tracing to model imaging data taken with the Mosaic II and BTC cameras. We have tentatively identified the dominant distortion patterns empirically measured by Jarvis & Jain (2005) with focusing errors (coupled to astigmatism in the primary mirror), misalignments or tilts between the primary mirror and the optical axis defined by the camera and corrector (inducing coma), guiding errors, and trefoil distortions of the primary mirror associated with its support system.

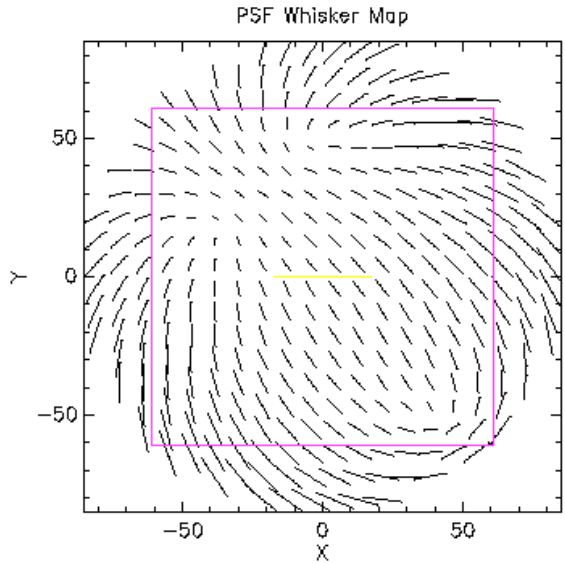

Figure 3.2. Predicted PSF pattern from an optical model of the existing Blanco corrector, with the primary mirror offset by .2 mm in x and -0.7 mm in y from the optical axis. In situ measurements have indicated the presence of time-dependent offsets of this magnitude, likely due to failure of some of the primary mirror radial supports. The inscribed box shows the area covered by the Mosaic II CCDs. The offset introduces coma in addition to that induced by the atmospheric dispersion compensator (ADC). This pattern is qualitatively similar to some PSF distortion patterns seen with the Mosaic II imager.

DECam is being designed and the telescope upgraded and enhanced to reduce or eliminate these systematic effects. The reference DECam optical corrector design achieves small and smoothly varying PSF distortions across the field of view, with an amplitude comparable to other recent or planned wide field optical correctors. DECam will be equipped with dedicated CCDs (absent from the Mosaic II), which will provide active control of the focus on the same cadence as the survey. The reference DECam optical corrector design does not include an atmospheric dispersion compensator (ADC), eliminating one source of coma. At present we are considering incorporating wave front sensors along with two-axis motions of the prime focus cage to actively recollimate the telescope during the night (see Section 8). The telescope is also being instrumented with position monitors to better understand the performance of the primary mirror support system and flexure of the telescope truss. Broken radial supports on the primary mirror have long been a source of difficulty, but the causes have been identified and a new design for attaching the rebalanced supports to the mirror is being studied and should resolve this problem.

*Tests for systematics:* Finally, the cosmological origin of the gravitational lensing signal can be tested from the DES data itself in multiple ways. We will develop the following set of tests using mock survey data. The goal will be to improve the analysis software and to establish precisely how well each will constrain the level of residual systematics:

1. *Tests for a cosmological signal:* measurement of the ellipticity correlations of stars; correlation of foreground galaxy shapes with background galaxy and quasar positions.

2. *Tests for gravitational shear:* E/B mode decomposition; three-point shear correlations for given amplitude of two-point signal.

3. *Calibration tests:* shear signal vs. galaxy size and surface brightness, in given redshift bins.

4. *Alternative lensing measurements:* Magnification bias induced cross-correlations in the number density of galaxies. Number counts behind clusters.

Our cosmological parameter analysis will include nuisance parameters for residual systematic errors of the three types described above (Huterer et al 2005).

**4. Galaxy Clustering Studies of Cosmology**

The Dark Energy Survey (DES) will deliver a sample of over 300 million galaxies with photometric redshifts. This sample will extend to z~1.4, and by coordinating our survey with a planned, large solid angle VISTA-IR survey, we can extend accurate galaxy redshift estimates up to z~2 and beyond (see Section 6). The 5000 deg$^2$ solid angle, the 300 million galaxies and the redshift depth make the DES galaxy sample extremely well suited for a galaxy clustering-based study of the dark energy.

On large scales, galaxy clustering and its evolution reflect the gravitational clustering and dynamics of the underlying dark matter distribution. The ratio of the galaxy and dark matter power spectra can be described by a redshift and scale dependent bias factor. This bias factor $b^2(z)$ is expected to be scale-independent on large scales, although its amplitude does depend on the type (e.g., luminosity, color) of galaxy being studied. Moreover, on sufficiently large scales the clustering is solidly in the linear regime ($\delta\rho/\rho\ll1$). In this linear regime, we can write the galaxy power spectrum as

$$P_{gal}(k) \propto k^n T^2(k;p_i)\, g^2(z;p_j)\, b^2(z),$$

where the initial dark matter power spectrum from the early Universe $\propto k^n$, $T(k)$ is the scale-dependent transfer function for dark matter perturbations, $g(z)$ is the scale-independent linear perturbation growth function, and the $p_i$ remind us that these functions depend explicitly on cosmological parameters. In practice, we replace this simple bias model with the more physically motivated and numerically justified halo occupation model mentioned above. From this equation it is clear that a measure of the galaxy clustering provides direct information about cosmology. Moreover, characteristic scales in the galaxy power spectrum, including the horizon scale at matter—radiation equality and the baryon oscillations, provide physically calibrated standard rods that can be used to measure the angular diameter distance as a function of redshift (e.g., Cooray et al 2001). Because this method primarily uses the shape of the clustering power spectrum rather than its amplitude, it is relatively insensitive to uncertainties in galaxy biasing.

An important consideration is the fact that our galaxy photometric redshifts will have characteristic errors of $\sigma(z)$~0.07 compared to typical spectroscopic redshift uncertainties of $\sigma(z)$~0.0001. Thus, the effective resolution of our survey along the line of sight is significantly degraded relative to the resolution perpendicular to the line of sight. This raises interesting questions about the optimal way of studying the galaxy clustering. One could carry out a 3 dimensional analysis of $P_{gal}(k)$ that accounts for the loss of information along the line of sight. Alternatively, one could study the angular correlation function or angular power spectrum within redshift shells. While the angular correlation function is well suited for studies of clustering properties on small scales, it is the angular power spectrum that is best suited for studying clustering on large angular scales where we expect the features with most cosmological information to reside. We will carry out an exhaustive study of the optimal technique for extracting information from this kind of dataset, but in the following discussion we focus on the galaxy angular power spectrum measured within redshift shells.

The angular power spectrum within a redshift shell can be written as

$$C_{gal}^i(l) = \int_0^\infty k^2 dk \frac{2}{\pi} f_i^2(l,k) P_{gal}(k),$$

where $f_i(l,k)$ is the Bessel transform of the radial selection function for redshift shell $i$ (Tegmark et al. 2002, Dodelson et al. 2002). The radial selection will be computed from the photometric limits of our survey along with the photometric redshift error distribution. We use a Fisher matrix analysis to estimate the resulting dark energy constraints; we marginalize over 5 halo parameters in each redshift shell, assume spatial flatness, and restrict information to angular multipoles $l < 300$ where the halo model of bias is most robust. We include CMB priors on the power spectrum, employ photometric redshift shells with $\delta z=0.1$ extending from $0.1<z<1$, and include galaxies in halos with mass greater than $10^{12.5} h^{-1}$ solar masses. Fully marginalized constraints on constant $w$ models are given in Table 1 of the White Paper.

*Systematic Errors:* There are several potential systematic effects that concern us: (1) scale dependent galaxy bias on the largest scales, (2) photometric zero point drift over the survey, and (3) photometric redshift biases. For the first, the concern is that although there is no known physics that would introduce scale dependencies in the galaxy bias on the ~100 Mpc scale, such a scale dependence could bias our dark energy measurements. One way of minimizing sensitivity to this effect is to use only the baryon oscillation features and exclude the more gradual break to large scales, which can be traced to the horizon scale at matter radiation equality. The constant $w$ constraints weaken by approximately a factor of two in this case.

For the second concern, the photometric zeropoint drift affects the number density of galaxies in each redshift bin that lie above the detection threshold. We calculate that a $\delta m_{zp}$~0.01 shift results in a fractional change in the number density $\delta n/n$ of detected galaxies that scales roughly linearly with redshift and reaches 0.7% at z=1. To calculate the effect this will have on our measurement, one must model the angular power spectrum of the photometric zeropoint drift. Assuming there are no preferred scales for the zeropoint drift, a drift at the level of the DES science requirement ($\delta m_{zp}$ =0.01) would introduce an irreducible, fractional noise contribution in the angular power spectrum $\delta C_l / C_l$~$(\delta n/n)^2$=1.4%. Including this noise source leads to an increase in the error of $\delta w<0.05$ (in a single redshift bin, Figure 4.1) on the constant $w$ constraint from baryon oscillations and $\delta w<0.03$ from the horizon scale at equality.

We can think of a photometric redshift bias as a bias in the measured distance-redshift relation. The DES science requirements drive us to control the photometric redshift biases at the level of $\delta z$~0.001 per photo-z bin, and Section 6 shows that this can be achieved with the expected spectroscopic calibration sample. This redshift error corresponds to $\delta w$~0.007 at z=0.5 and $\delta w$=0.001 at z=1 for the angular clustering method. Thus, photometric redshift biases would have to be more than an order of magnitude larger to become significant effects in our forecasts.

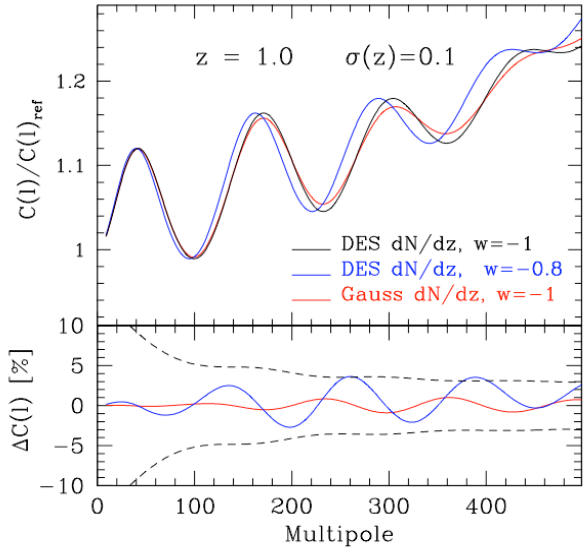

Figure 4.1. Top panel shows predictions for the baryonic oscillations from non-linear angular power spectrum $C_l$ in the DES for w=-1.0 (black) and w=-0.8 (blue) in a redshift slice of thickness Δz=0.1 at z=1. Red line shows the degradation of the w=-1 case with 0.1 rms photometric errors. Bottom panel shows relative cosmic errors (dashed line) compared to relative variations in the lines above. Cosmic errors give δw=0.20 (for a single slice) while photometric errors correspond to δw=0.05.

While our main focus for probing dark energy is using the shape of the $C_l$'s we will also carry out a more complete modeling of the galaxy clustering that includes the full amplitude information for the power spectrum and the angular bispectrum. Cosmic errors on the amplitude are smaller than the corresponding errors on the shape, but the systematic uncertainties are quite different. This approach requires modeling galaxy bias and a more accurate control over the galaxy selection function, which mostly affects the amplitude of clustering. On large scales, galaxy bias can be self-calibrated to a few percent accuracy using the reduced bispectrum (see, e.g, Sefusatti & Scoccimarro 2005). On smaller scales biasing can be modeled with the halo occupation distribution. This modeling can be constrained with the clustering as a function of galaxy luminosity and color. This will provide important clues to galaxy formation. Marginalizing over the shape information, the amplitude of clustering on large scales constrains the linear growth function in a way that is independent of the modeling of the primordial spectrum or the matter transfer function. This will provide complimentary information on the dark energy equation of state with very different systematics. The details of this modeling are still in progress, but given the wide range of redshifts covered by the DES, there is a great potential in this approach.

## 5. Type Ia Supernovae

Using the information contained in supernova (SN) light curves to measure the expansion history of the universe has rapidly become a foundational standard of cosmological studies. Studies of nearby SNe provided the basis for development of methods of using Type Ia SNe as precision distance indicators based on their maximum apparent magnitudes and decline rates or "stretch" factors (*e.g.,* Hamuy et al. 1996, Riess, Press, & Kirshner 1996, Perlmutter et al. 1997), and the application of these methods to studies of high redshift SNe provided the first direct evidence of the accelerating expansion of the Universe (Riess et al. 1998, Perlmutter et al. 1999). While most of the success of these studies in constraining the cosmological expansion based on cosmological model-independent luminosity distances has been due to the precision to which the light curves can be measured and compared against local templates, the basic **cosmological systematic uncertainty** in the analysis lies in the resulting elongated (banana) contours in the $\Omega_m$-*w* plane. Much of the power in determination of cosmological parameters relies upon the complementary nature of the confidence contours derived from SN studies with those derived from studies of galaxies and large scale structure, such as the other studies planned with the DES.

SN surveys subsequent to those which identified the evidence of the accelerating universe have focused on minimizing or eliminating systematic uncertainties using new measurement capabilities and larger samples. The DES offers the opportunity to make the next step forward in this progression. Compared to the current generation of supernova surveys (*e.g.*, ESSENCE and CFHT SN Legacy Survey), we will have a significantly wider field to collect larger numbers of supernovae over a wide range of redshifts. The DECam design, together with the advanced calibration system currently being developed for the Blanco 4m telescope (Stubbs et al. 2005), will allow much better understanding and control of the wavelength response of the entire photometric system, and the proposed fully-depleted CCD sensors will allow much better throughput in the redder wavelengths, particularly *z* band which is crucial to cover the desired range in redshift, out to z~0.75. Together, the optics, detectors, and calibration system allow us to minimize the **instrumental systematic uncertainties**, providing better photometric precision than previous surveys as well as better calculation of the K-corrections, which are key to comparing the light curves of the higher redshift SNe with the local templates.

Based on these new capabilities, we have designed a baseline SN experiment which uses approximately 10% of the time dedicated to the DES operations, assumed to be a third of the telescope time over a five year period. The requirements of this design include the production of a large number of well-sampled SN light curves in three bands in an observing strategy which fits within the 5000 deg$^2$ DES survey area and survey strategy. Based on previous modeling, balancing desired spatial coverage with desired depth to cover a wide range of redshifts (0.25 < z < 0.75), we have selected exposure times of 200s in *r*, 400s in *i*, and 400s in *z*. These exposure times should give us reasonable signal to noise light curves in these bands for SNe out to z~0.75.

The 40 deg$^2$ SN survey will use a repetitive six day cycle to observe every field in *r* twice and every field in *i* and *z* once. The *r* images will detect SNe and measure their light curves. The *i* and *z* band images will provide color information for SN typing and determination of SN photometric redshifts. Stacked images will provide very deep photometry of the host galaxies. Field placement will depend on the geography of spectroscopic follow-up resources, but the current plan is to choose several of the fields to overlap the deep spectroscopic galaxy surveys of ~16 deg$^2$ being carried out by the VLT. This will ensure that a reasonable fraction (~25%) of the detected SNe Ia lie in galaxies with previously measured spectroscopic redshifts and will calibrate SN photo-z's. SN and galaxy photometric redshifts will be used for the remainder of the sample with neither host nor SN spectroscopy. The area covered in the SN survey will be much larger than that covered by any current intermediate to high redshift SN surveys.

With this baseline design, we have run Monte Carlo models assuming that the DECam has roughly similar *r* and *i* response to that of the current CTIO Mosaic II camera (a conservative assumption) and with the improved *z* response described in Section 9. Folding these sensitivities in with the historical weather, seeing, and other observational factors, we estimate that we will identify more than 2000 Type Ia SNe (along with many SNe of other types) over the course of the five year program.

Using this large sample of well-characterized SNe, we can further constrain cosmological parameters such as *w* (beyond the constraints planned by current surveys) with the whole sample while investigating **astrophysical systematic uncertainties** by comparing selected subsamples (such as subsamples at similar redshifts or with similar decline rates). We have propagated the simulated light curves through a sample analysis and shown that we can constrain *w* to ~10% (4%) using WMAP (Planck) priors and using a broad range of simplifying assumptions, including knowledge of the SN redshifts and accurate calibration of the observing system to minimize uncertainties in K corrections. These constraints are shown in Figure 5.1, which demonstrates the complementarities between SN and CMB parameter constraints.

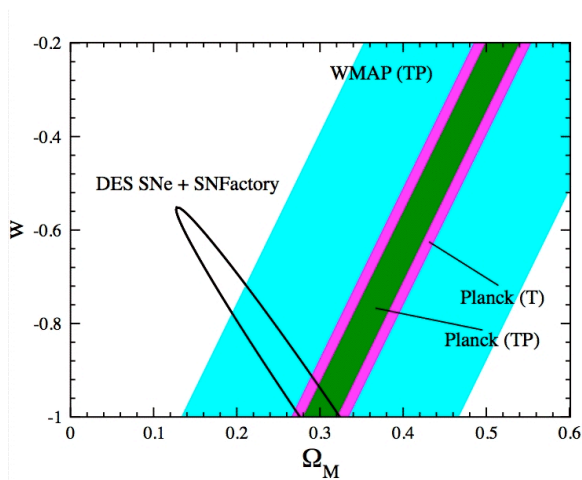

Figure 5.1 DES SN constraints and CMB constraints on the dark energy equation of state and dark matter density, assuming a flat Universe. 'SNFactory' here denotes that a sample of ~300 nearby SNe Ia are assumed to anchor the Hubble diagram at low redshift.

Spectroscopy of all of the SNe we discover (roughly 100 Type Ia SNe per month, as well as many SNe of other types) is unlikely to be feasible, even with the combined efforts and 8m-class telescope time of the extended DES collaboration. We can, however, rely upon the photometric redshifts of the host galaxies generated by the DES, as well as complementary spectroscopic measurements provided by other surveys which overlap the area covered in the DES SN survey, such as the VIMOS-VLT Deep Survey (VVDS) and the NOAO Deep Wide Field Survey (NDWFS). We estimate that we should have photo-$z$ estimates for more than 80% of the host galaxies of the SNe we discover and be able to obtain spectroscopic redshifts for roughly 20%. Our preliminary studies indicate that the errors in host galaxy photo-$z$ are anticorrelated with both the K-correction errors and the uncertainties in the decline rate or width versus brightness relations. Based on these estimates, we can predict that the RMS of our SN distance determinations may increase from the ~0.15 magnitudes typical of well studied samples to ~0.25 magnitudes with host-galaxy photometric redshifts only. In this case, the statistical errors on $w$ increase to 0.14 for WMAP prior but are virtually unchanged with the Planck prior. We note that if we adopt a crude model for systematic errors, with a floor of $0.02(1+z)/1.8$ mag in the dispersion of SN peak magnitudes in redshift bins of width 0.1 (added in quadrature to the intrinsic dispersion of 0.25 mag), then the cosmological parameter errors degrade only modestly to 0.15 for WMAP and 0.04 for Planck; these are the values quoted in Table 1 of the White Paper.

In addition to the photometric redshifts of host galaxies, we will rely upon photometric redshifts measured from the SNe themselves. While this practice is relatively new, several groups have made significant progress over only the last year or two, including Barris & Tonry (2003), Howell et al. from the SNLS (in prep.), Prieto et al. from ESSENCE (in prep.), and Pinto (private communication). Figure 5.2 shows a sample high-redshift light curve from ESSENCE for which the photometric redshift was accurately estimated using these techniques. This estimate was made without any prior based on the photometric redshift of the host galaxy. Although much work is still needed, the combination of photometric redshifts from the host galaxies and the SNe themselves should provide the information necessary to derive cosmological constraints from the sample of SNe obtained over the course of the DES. Using the foundation of the DES simulations described elsewhere, we plan to explore the accuracy we can achieve and the possible systematic errors which could be lurking in the application of these new methods. Preliminary studies indicate that a SN photo-$z$ redshift accuracy of 0.02 is sufficient to keep the SN constraints on $w$ from degrading by as much as 20% (Huterer et al. 2004).

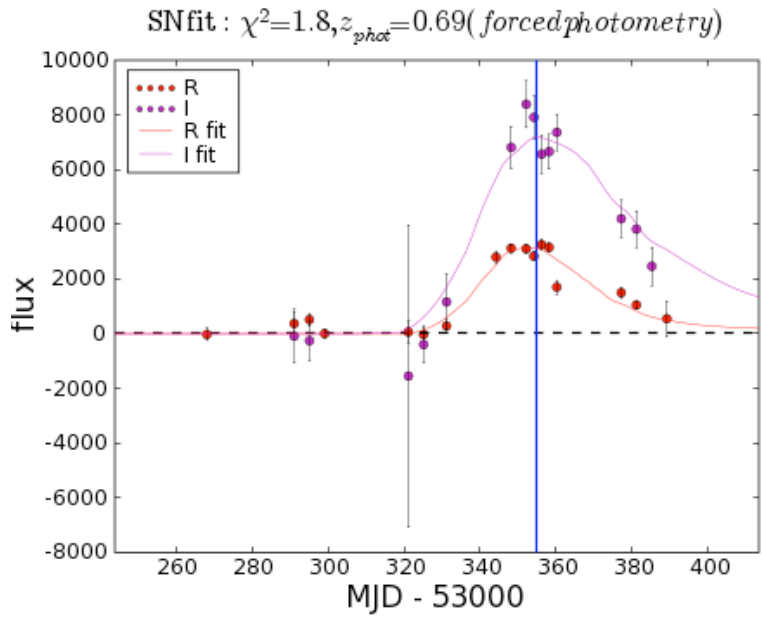

Figure 5.2. Sample light curve fit from the ESSENCE SN survey based on the real-time photometry (not final photometry, which should be significantly better). The photometric redshift based only on the observed R and I light curves was calculated to be z=0.69, while the spectroscopic redshift derived from the host galaxy was measured to be z=0.68.

It is important to note that the DECam would be one of the primary instruments in the world for this type of supernova survey during its years of operation. We expect that the current generation of SN surveys will yield suggestive results and puzzles that will be addressed in detail by the DES SN work, so the baseline plan described above is very likely to be updated with a significantly more sophisticated plan that builds on what we know at the time we begin this effort. It is also significant that many of the challenges we face in the DES SN survey, such as the lack of spectroscopic information on the majority of SNe, are the same as those which will be faced in the future when we attempt to use the incredible samples of SNe which LSST will identify continuously throughout its operational lifetime. Just as the results of the current generation of SN surveys will guide us in updating and refining the strategies we will employ for the DES SN survey, so will the DES provide crucial experience and guidance for deriving scientific results from the plethora of SNe discovered by LSST.

## 6. Photometric Redshifts

In order to achieve its scientific goals, the Dark Energy Survey will need to obtain accurate galaxy photometric redshifts (photo-z's) derived from the DES *griz* imaging data. Detailed understanding of the photo-z error distributions as functions of galaxy magnitude, redshift, and type, will be important for obtaining accurate cosmological parameter constraints. In particular, very large completed and ongoing spectroscopic redshift surveys will be available before the start of DES observations, and they will provide the data sets we need for accurate calibration and measurement of photo-z's and photo-z errors, down to the DES photometric limit. Below we will begin by describing the results of our simulation efforts to characterize and optimize the photo-z quality expected for DES galaxies and galaxy clusters. Next, we will discuss the requirements on how well we need to understand our photo-z error distributions, so as not to compromise DES cosmological parameter constraints. Finally, we will describe the large spectroscopic redshift training sets we will use to meet our requirements on accurate photo-z calibrations for DES.

The DES cluster key project requires photometric redshift measurements for cluster galaxies, and such photo-z's are greatly facilitated by the strength of the 4000Å break feature prominently seen in the spectra of red cluster galaxies. We use Monte Carlo simulations to assess the quality of DES cluster photometric redshifts. We adopt the local cluster luminosity function and luminosity-mass and number-mass relations of Lin, Mohr, & Stanford (2004), and our cluster galaxies follow a passively evolving elliptical galaxy model from the Pegase-2 library (Fioc & Rocca-Volmerange 1997), for a flat cosmology with $\Omega_M=0.3$, $\Omega_\Lambda=0.7$, and $h=0.7$. The cluster luminosity function faint-end slope is fixed at $\alpha = -1.1$, and we take the halo occupation number to evolve with redshift as $(1+z)^\gamma$, with $\gamma=1$ (Lin et al. 2004; Kravtsov et al. 2004). We use the DES $10\sigma$ $griz$ galaxy magnitude limits and add in quadrature a 2% error from photometric calibration uncertainty. A least-squares template fitting method is used to determine photometric redshifts for clusters with $1.0 \times 10^{14}$ and $2.5 \times 10^{14}$ solar masses, and in each case 20,000 mock clusters are generated and distributed uniformly over the redshift range z=0–2. Fig. 6.1 shows our results and demonstrates that the DES will provide robust photometric redshifts for such clusters to $z \approx 1.3$. For these clusters, we find a small photo-z scatter (68% limit) $\sigma_z \leq 0.02$, with the tails of the photo-z error distribution extending no more than about 0.05 in redshift. At higher redshifts, $z > 1.3$, color degeneracies become important, and the tails of the error distribution become larger, though the 68% limit scatter is still typically $\sigma_z < 0.1$.

The DES weak lensing, angular power spectrum, and supernova projects will also require photo-z measurements for the general field galaxy population. Such photo-z's are necessarily less accurate than those possible for cluster galaxies, as we must consider a much broader distribution of galaxy types. Nonetheless, our simulations show that the DES will obtain well-behaved photo-z's, with scatter (68% limit) $\sigma_z < 0.1$, out to redshifts $z > 1$ (Cunha et al. 2005). For our Monte Carlo simulations, we adopt the galaxy magnitude-redshift distribution derived from the luminosity function results of Lin et al. (1999) and Poli et al. (2003), combined with the galaxy type distribution derived using data in the GOODS/HDF-N field (Capak et al. 2004, Wirth et al. 2004, Cowie et al. 2004). We simulate a flux-limited sample of 100,000 galaxies, with redshifts $0 < z < 2$, magnitudes $20 < i < 24$, and compute photometric errors according to the DES 5-year $10\sigma$ $griz$ magnitude limits. To optimize our photo-z's, we tested several different techniques, specifically polynomial fitting (e.g., Connolly et al. 1995), neural networks (e.g., Collister & Lahav 2004), and template fitting (e.g., Bolzonella et al. 2000). As shown in Fig. 9, we find that our best results are derived from a hybrid "comparison" technique, where we average the photo-z's derived from the neural network and template-fitting methods, obtaining an overall scatter (68% limit) $\sigma_z = 0.07$, as well as reasonably well-behaved photo-z error distributions. In addition, we have examined the impact of adding near-IR filters, specifically 15 minute exposures on VISTA, with $10\sigma$ limits (AB mags.; from W. Sutherland) Y=22.45, J=22.15, H=21.65, Ks=21.15. As Fig. 6.3 shows, we find a reduced overall scatter (68% limit) $\sigma_z = 0.05$, as well as significant (and expected) improvement at $z > 1.4$, where the critical 4000Å break redshifts out of the DES $z$-band filter. Note that very similar results are obtained with just a single VISTA filter if it is J or redder. We are planning to further characterize the expected DES photo-z error distributions for both cluster and field galaxies, using improved mock galaxy catalogs drawn from large N-body simulations (see Section 7).

A central goal of the DES photo-z effort is the detailed characterization of the photo-z error distribution as a function of magnitude, redshift, and galaxy type, since accurate cosmological parameter estimation for each of the 4 projects requires accurate knowledge of the photo-z error distribution. Note in particular that the limiting systematic error in degrading the cosmological parameter constraints is not the absolute size of any photo-z bias or scatter, but rather the *uncertainty* in knowing what that bias or scatter is. Specifically, we can divide the DES galaxy or cluster sample into photo-z bins and examine the effect on the cosmology constraints due to uncertainties in the photo-z bias and scatter in those bins (Ma et al. 2005, Huterer et al. 2004). This is illustrated in Fig. 11 for shear-shear tomography, which shows that we need the photo-z bias uncertainty to be < 0.002 and the scatter uncertainty to be < 0.01 in each redshift bin in order to have <10% degradation in the constraint on $w$. A similar analysis for the cluster method indicates that we need an accuracy of 0.001-0.005 in the cluster photo-z bias in bins of width $\Delta z=0.1$; this is shown in Fig. 6.5. We are actively working to further quantify these photo-z requirements and to extend the analysis to the DES angular power spectrum and supernova projects.

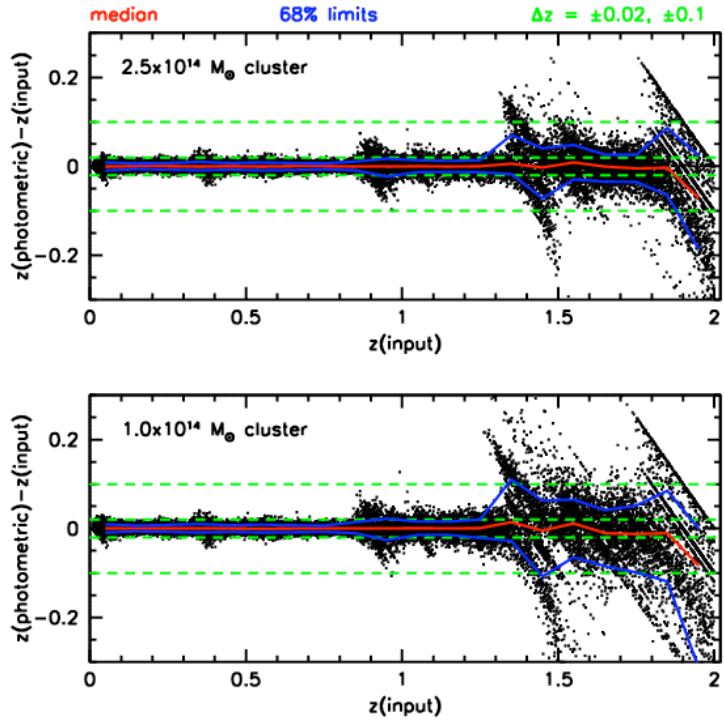

Figure 6.1. Photometric redshift results for Monte Carlo simulations of $1.0 \times 10^{14}$ and $2.5 \times 10^{14}$ solar mass galaxy clusters; see text for details. The red lines show the median difference between photometric and true redshift, the blue lines show the scatter (68% limits), and the green dashed lines are set at $\Delta z = \pm 0.02$ and $\pm 0.1$ to guide the eye.

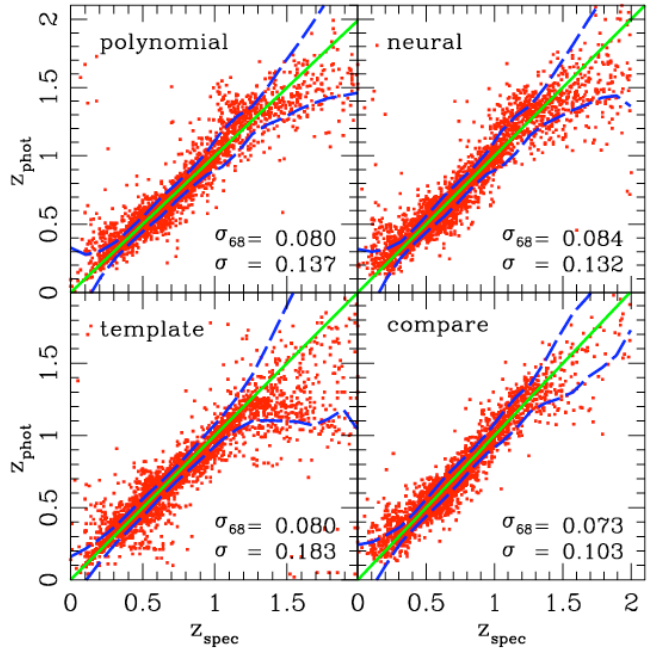

Figure 6.2. Photometric redshift results for Monte Carlo simulations of the general field galaxy population for DES; see text for details. The green lines indicate where photometric redshift = true redshift, and the blue dashed lines show the 68% limits. $\sigma$ is the overall rms photo-z scatter, while $\pm\sigma_{68}$ indicates the 68% limits. The panels show the results using different photo-z techniques: polynomial fitting, neural networks, template fitting, and a hybrid "comparison" method (average of neural network and template fitting photo-z's) which gives the best results.

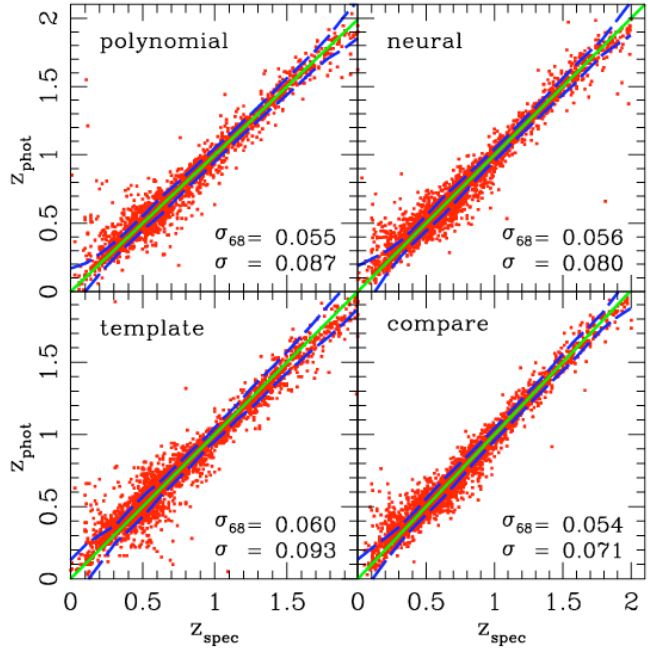

Figure 6.3. Same as Fig. 6.2, but now adding YJHKs near-IR data from VISTA to the original DES griz data to improve the photometric redshifts; see text for details.

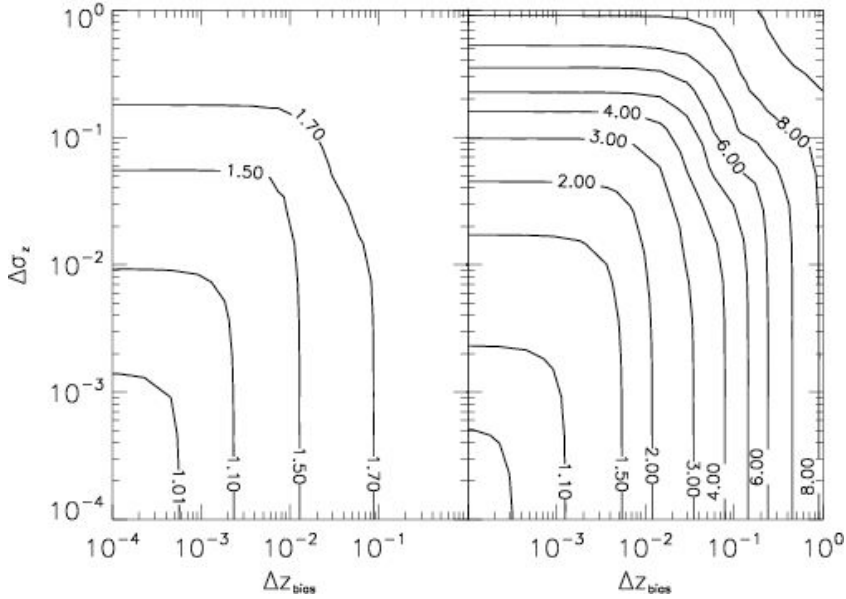

Figure 6.4. The degradation in dark energy cosmological parameter constraints, derived from weak lensing shear-shear tomography using photo-z bins, for a given error in assumed photo-z bias (x-axis) and photo-z scatter (y-axis) in each bin. Seven photo-z bins are used in the redshift range $z = 0-2$. The left panel assumes a constant-$w$ model, while the right panel allows $w$ to vary. The contour levels indicate the amount of degradation of the parameter constraints, e.g., in the left panel the constraint on $w$ weakens by a factor of less than 1.10 if the error in photo-z bias is < 0.002 and the error in photo-z scatter is < 0.01.

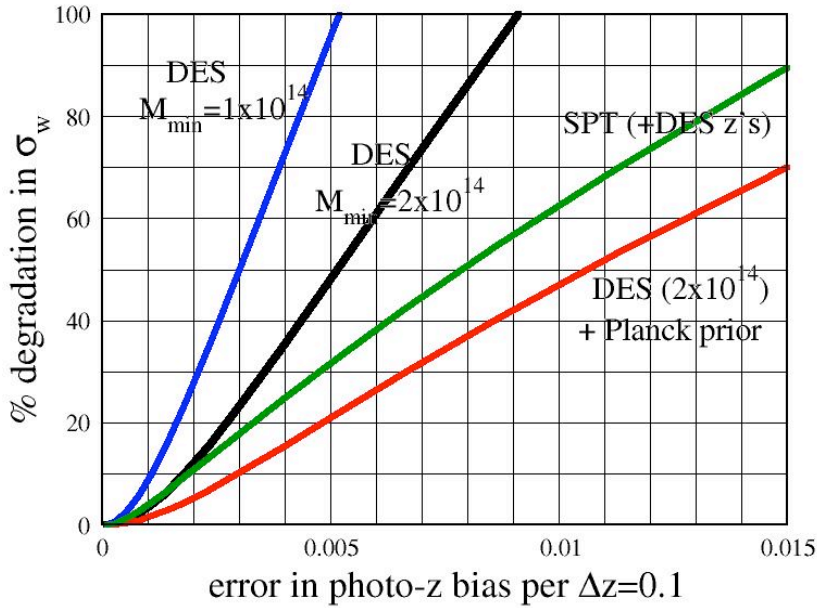

Figure 6.5. For DES cluster counting, the degradation of the measurement uncertainty $\sigma_w$ on the dark energy equation of state parameter $w$ is shown as a function of the error in the photo-z bias for bins of width $\Delta z=0.1$. The curves illustrate the cases of fixed minimum cluster mass limits of $1\times10^{14}$ and $2\times10^{14}$ solar masses (the latter with and without Planck priors) and the case of clusters detected above the SZE flux limit of the SPT.

These photo-z requirements will necessitate use of training sets of $5 \times 10^4 - 10^5$ spectroscopic redshifts, and we will rely on a number of ongoing or completed redshift surveys to provide the large galaxy samples we need for accurate calibration of DES photometric redshifts. At bright magnitudes, the DES will rely primarily on the Sloan Digital Sky Survey (SDSS) and on the 2dF Galaxy Redshift Survey (2dFGRS). The DES area intentionally covers the SDSS Southern Equatorial Stripe, which not only includes standard SDSS main galaxy and luminous red galaxy spectroscopic redshifts but also various deeper SDSS Southern Survey spectroscopic samples extending down to $r \approx 20$; in total, some 70,000 SDSS spectroscopic redshifts lie within the DES area. Likewise, the DES will overlap significantly with the 2dFGRS area, making another 90,000 spectroscopic redshifts available to DES at bright magnitudes $b_J$ < 19.45. At faint magnitudes, down to the DES limit of $i \sim 24$, we will use two ongoing deep redshift surveys: the VIMOS VLT Deep Survey (VVDS; Le Fèvre et al. 2005) and the Keck DEEP2 Survey (Davis et al. 2004). The DES will cover 3 VVDS fields (including both the VVDS equatorial deep field and the Chandra Deep Field South), which will provide about 60,000 spectroscopic redshifts down to $I_{AB}$ = 24. The DES will also cover both of the DEEP2 fields on the Southern equator, which should provide about 30,000 spectroscopic redshifts down to $R_{AB}$ = 24.1. In addition, we will repeatedly image these deep redshift survey fields as part of the DES supernova survey, so that we will obtain deep and well-calibrated photometric data for these faint training set galaxies. We will also carefully examine issues of sample completeness and fairness for these training sets, in order to identify any potential regions of redshift or galaxy parameter space with more uncertain photo-z's, so as to enable us to use only well-understood photo-z galaxy samples in our science analyses. Overall, the availability of some 250,000 spectroscopic redshifts before the start of DES observations will provide us with the necessary training sets to optimize our photo-z techniques, accurately characterize photo-z error distributions, and sufficiently control our photo-z systematic errors so as not to compromise our cosmological parameter constraints.

## 7. Simulations

We envision producing a series of catalog and image simulations for the Dark Energy Survey, for the purpose of helping us to develop our science analysis codes and data reduction pipelines prior to the start of survey observations in 2009. Moreover, the simulations will allow us to validate our analytical forecasts of dark energy parameter constraints, as well as to extend our analyses to comprehensively characterize and incorporate the various sources of statistical and especially systematic errors inherent in each of our 4 project techniques. We plan to have yearly cycles of new catalog and image simulations, followed by science analysis and data reduction challenges carried out using the simulation outputs. The level of scale and sophistication involved in each round of simulations will improve, in order to meet the requirements set in conjunction with the science analysis goals of the key project science working groups, and with the data reduction/pipeline development testing goals of the data management project. We outline below our current simulation plans and briefly describe each level of DES simulations.

<u>Level 0 Image Simulations (2004):</u> These image simulations (ImSim0) were intended to provide an early jump start for the data management effort to test processing pipelines. We used existing reduced SDSS southern equatorial coadded data, remapped the images to the DECam plate scale, and applied atmosphere, optics, and detector models to simulate raw, unprocessed DES data. In total 55 DES fields (165 square degrees overall) were simulated, each with *griz* images, for a total of 220 GB of data. The images were generated at Fermilab, transferred to and processed at U. Illinois and NOAO using two independent pipelines, and the output object catalogs were successfully validated against catalogs based on the original input SDSS images.

<u>Level 1 Catalog and Image Simulations (2005):</u> For the level 1 DES simulations, we are producing a 400 square-degree catalog of galaxies and stars (CatSim1), followed by a 300 square-degree subset with images (based on CatSim1 outputs), corresponding to about 1 week of DES imaging data (ImSim1). We are using this first round to begin to develop our simulation tools, and our basic science analysis goals include: recovery of input object magnitudes, shape parameters, and shears; tests of photometric redshift measurement and cluster finding algorithms; and recovery of cluster-shear correlations and angular power spectra. The data reduction challenge goals include tests of the data processing framework, the astronomy modules, and the image data archive.

CatSim1, the level 1 catalog simulation, consists of 4 components:
- Galaxy Catalog: Mock galaxies are assigned to the dark matter particles from the Hubble Volume Simulation (with $\Omega_M=0.3$, $\Omega_\Lambda=0.7$, $h=0.7$, $\sigma_8=0.9$, and redshift z<1.4), according to the SDSS r-band luminosity function (plus passive evolution) and using the ADDGALS method (R. Wechsler), which matches the observed luminosity-dependent 2-point correlation functions from the SDSS. Each mock galaxy is assigned the properties of a real SDSS galaxy, based on its luminosity and local density, so that the procedure reproduces the local luminosity-color-density correlations from the SDSS.
- Shapelets Catalog: Each mock galaxy is assigned shapelet coefficients based on its corresponding SDSS galaxy.
- Shear Catalog: A gridded shear field (to $l\sim10^4$) is determined using the Born approximation from the dark matter distribution in the parent Hubble Volume Simulation, and then interpolated to the position of each mock galaxy.
- Stars Catalog: Bright stars with r<18 are taken from the real USNO-B catalog over the simulation area. Faint stars with r>18 are drawn from the Besancon stellar population synthesis model for the Milky Way.

ImSim1, the level 1 image simulation, will be generated using the following tools:
- Observing and atmosphere model: The survey observing cadence is simulated, along with the atmospheric conditions, including transparency, extinction, seeing, and sky brightness.
- Optics model: The telescope and corrector optics are modeled, and in particular the griz PSFs due to the optics will be tabulated as a function of focal plane position and telescope focus.
- Instrument model: The CCD detector characteristics (QE, read noise, gain, cosmetics, etc.) will be taken from actual LBNL devices.
- Astronomical objects renderer: A java-based tool has been developed which renders galaxies and stars using shapelets (from CatSim1), and adds noise and convolves/transforms objects in accord with the atmosphere, optics, and detectors

Level 2 Catalog and Image Simulations (2006): Here we will produce a 5000 square-degree catalog (CatSim2) covering the full DES area, plus images amounting to 1 month of DES data (about 1200 square degrees; ImSim2). Our main science analysis goal is the recovery of the input cosmological parameters using each of the 4 project methods at the catalog level. We will build upon our level 1 simulation tools and make a number of improvements, in particular a new catalog component consisting of supernovae with lightcurves. Moreover, we will need to increase the dark matter simulation resolution to properly simulate low-mass halos and low-luminosity galaxies, while still maintaining a large box size. For CatSim2 we are aiming for a mass resolution of $\sim10^{10}$ solar masses, significantly better than the $\sim10^{12}$ solar mass resolution (corresponding to M* halos) of the Hubble Volume Simulation used in CatSim1. Ray tracing will also be needed to generate accurate weak-lensing shear fields. In addition, we will improve our modeling of galaxy properties (luminosities, colors, clustering, shapes, etc.) and their redshift evolution, using deeper data sets including RCS/CNOC2, ALHAMBRA, and DEEP2. Besides the primary catalog and image simulations, we also plan to carry out more specialized simulations to address various issues for the key project analyses. These include N-body simulations with hydrodynamics that would be used to calibrate the theoretical cluster mass function, to improve models of SZ-cluster selection, and to understand correlations between optical and SZ clusters. Likewise, we plan simulations to investigate the effects of small-scale nonlinear clustering and of baryons on the weak lensing and galaxy angular clustering analyses. Finally, for weak lensing we will also improve our image simulation tools to permit modeling of various systematic effects, in particular temporal and spatial correlations of the atmospheric PSF, telescope optical alignment and related issues, and shear calibration. To carry out our simulations program, we will draw upon the computing resources and simulations expertise within the DES collaboration.

Level 3 Catalog and Image Simulations (2007-8): Here we will produce a suite of full-DES catalog simulations (CatSim3) with different input cosmologies, plus a single set of image simulations for 1 year of DES data (5000 square degrees, 2 tilings; ImSim3). The image simulations will be used by the data management project to drive a full stress test of the entire data processing system. Our science analysis goals will include recovery of the input cosmological parameters from the suite of catalog simulations, as well as from the pipeline-reduced catalog outputs of the image simulations. In addition, as part of our simulation effort, we will continue to model systematic errors and related issues as needed for the 4 key project analyses, as well address some secondary DES science goals, including galaxy-CMB cross-correlations (e.g., ISW effect), galaxy formation and evolution, galaxy cluster physics, and quasars.

**8. Blanco Telescope**

The V. M. Blanco 4-m telescope is sited at Cerro Tololo Inter-American Observatory – one of the world's premier observing sites. It is a wide-field Ritchey-Chrétien design, the f/2.7 prime focus has been exploited over the past two decades with a series of wide field CCD imagers and optical correctors – the combination of which has made the facility the southern hemisphere's (and at times the world's) leader in optical imaging as gauged by the metric $\varepsilon A \Omega$ (efficiency × aperture × detector area). The critical observations leading to the discovery of dark energy were made with this telescope, and several other important surveys have been completed or are underway using the prime focus 8K×8K Mosaic II Imager. This instrument is presently scheduled for over 60% of the telescope time.

At its time of construction 30 years ago, the quality of the Blanco telescope primary mirror defined the state of the art and the surface quality has only been exceeded in the last decade with the advent of super-polishing. The enclosed energy as a function of diameter as measured during the testing of the Blanco primary at the time of its acceptance is given below:

| Enclosed Energy | Diameter in arcsec |
|---|---|
| 57% | 0.15 |
| 80% | 0.25 |
| 94% | 0.40 |
| 99% | 0.50 |

The Cerro Tololo observing site has a median image quality of 0.65 arc seconds. The site seeing together with the primary mirror quality will allow the Blanco together with DECam to deliver excellent quality images, limited primarily by atmospheric seeing.

The median delivered image quality, corrected to V-band and zenith, for long exposures with the prime focus Mosaic Imager is 0.9 – 1.0 arcseconds, and the DECam-Blanco combination will do better. This performance is the result of substantial improvements that have been made to the telescope and its environment, most of these in a series of upgrades that were initiated a decade ago. In brief, the telescope thermal environment was improved by large ventilation doors at the telescope level to promote more efficient flushing of the dome, by several air ventilation subsystems designed to optimize the temperature of the primary mirror and telescope structure, by covering the telescope dome with thermally insulated aluminum panels, by attending to thermal management throughout the building, and by installing a servo-control to optimize the system. At the same time, active optical control of the primary and secondary mirrors was introduced so that the intrinsic quality of the telescope optics could be effectively utilized; however closing the loop was possible only at the f/8 focus, where a wave front sensor was installed.

Although all Dark Energy Survey simulations have used current performance characteristics, the Cerro Tololo observing site itself has median image quality of 0.65 arc seconds, and thus there is still room to improve delivered image quality. Many improvements will be made through DECam itself, but a number of additional upgrades and enhancements will be applied to the telescope and dome in order to maximize DECam's potential:

(i) The primary mirror active-optics control loop will be closed using wave-front sensor CCDs integrated into the DECam focal plane. The method used will probably be a variant of curvature sensing (the "donut" method) now under development at CTIO for the SOAR telescope.This will improve the focus and point-spread function stability.
(ii) Metrology of the prime focus mirror as a function of orientation has shown that the mirror moves in its cell by an amount sufficient to introduce detectable coma. The probable cause has been identified and a definitive fix is in development to be installed later this year.
(iii) Relative motions of the prime focus cage and primary mirror due to telescope flexure will be measured and checked against an FEA model to estimate the need for active compensation to be built into DECam to further improve optical performance.
(iv) The existing Mosaic imager has no overall thermal control. A full thermal management model for DECam will be developed, and control equipment will be implemented, including glycol lines and air ducts as necessary to make the instrument as thermally neutral as possible. Cooled environments for instrument electronics and for off-telescope equipment such as a cryocooler compressor will be provided.
(v) The existing environment control system will be evaluated, and options developed for improving the thermal properties of the telescope and dome. A cost-benefit analysis will decide which of the proposed options will be implemented.
(vi) The telescope control system (TCS) will be upgraded or replaced, prior to the delivery of DECam. This will improve reliability through replacement of aging systems and improve performance in areas critical to the Dark Energy Survey – in particular, investigations have shown that it is feasible to drive the telescope the required 2 degree step between DES fields within the readout time of the camera (~17s) so that telescope slews will not limit survey efficiency.
(vii) The existing prime-focus optical corrector is thought not to achieve its designed optical performance, particularly as one element is damaged. DECam will incorporate a new optical corrector with modern design features not previously available, including SolGel coatings.
(viii) The existing Mosaic imager is refocused by interrupting observations and defocus limits delivered image quality in good observing conditions. DECam will incorporate contemporaneous focus control on the same cadence as the survey imaging, eliminating these issues.
(ix) DECam's larger imager area, faster readout (Mosaic II: 100s, DECam: <20s), better observing tools and real-time quality control will radically improve imaging observations at CTIO for the entire community.

## 9. The 2004-2006 DECam R&D Program

The primary goals for the 2004-2006 R&D period are focused in three main areas: CCDs, optics, and readout electronics. For the CCDs, we will establish the yield; learn to test and characterize CCDs; and to demonstrate packaging that meets the optical and electrical requirements of the survey. For the optics, we will finalize design and develop a firm cost and schedule estimate. For the CCD readout electronics, we will develop CCD testing systems and use these to prototype the prime focus cage camera readout system and components. Below we discuss these three efforts in more detail. During this period we are also developing the focal plane cooling system, the thermal control and packaging of the front end electronics, shutter and filter changing mechanisms, and the integration of the DECam data acquisition and controls with the systems at the Blanco. These efforts are going well and will not be included in the discussions below.

Procurement of quality CCDs has historically been the most difficult part of the construction of a focal plane. For DES we require 62 science-grade devices plus spares and this presents the most significant challenge to the project. To meet the science goals of imaging objects at high redshifts in a reasonable amount of time, we have chosen to use 250 micron thick, fully depleted devices as developed by LBNL. As shown in Figure 9.1, these devices are roughly a factor of ~13 more sensitive in the *z* band than the SITE devices in the existing Mosaic II Camera. Other sources were investigated but did not meet our performance, price, nor schedule requirements.

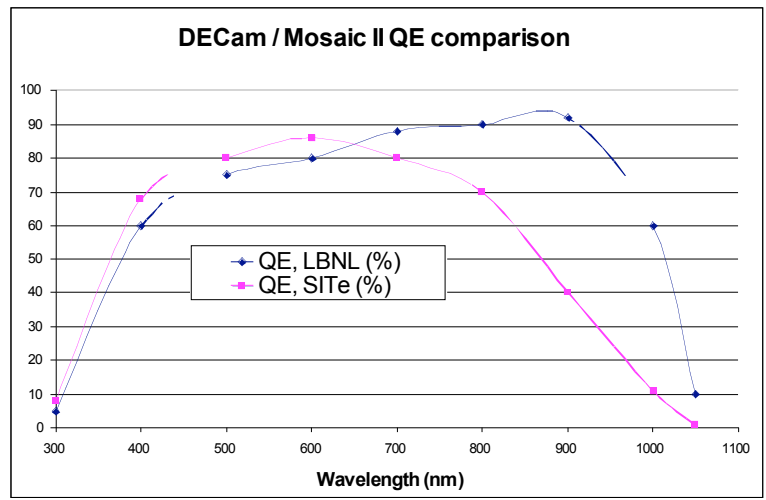

Figure 9.1. Quantum Efficiency of LBNL CCDs compared to those in the Mosaic II camera currently on the Blanco telescope.

LBNL has been developing CCDs in their Micron Systems Lab (MSL) since the early 1990's. Thick, fully-depleted CCDs fabricated at LBNL have been deployed on instruments at several ground based telescopes, including the Multi-Aperture Red Spectrograph at Kitt Peak and the Hamilton Echelle Spectrograph at Mt. Hamilton. They have enabled science in several areas including planet hunting, spectroscopy of very distant quasars, Seyfert galaxy black holes, and Type Ia supernovae. Most recently the effort at LBNL has been focused on development of devices for the SNAP/JDEM project. The device format appropriate for DES is a 2k x 4kpixels each 15 microns square. CCDs in this format have already been fabricated by LBNL and are currently being packaged at Lick Observatory for use on the LRIS spectrograph on Keck.

LBNL has developed a business relationship with a commercial vendor, DALSA, to reduce the processing load on the MSL and to increase production throughput for SNAP. In the SNAP business model, DALSA, performs 8 out of the 11 mask steps. The automatic processing equipment at DALSA is unable to handle devices less than 650 microns thick. A small number of wafers (3-4) are completely processed as thick devices for quality control. LBNL receives the wafers and subcontracts mechanical thinning to 250 microns thickness. LBNL then performs the last three mask steps and applies the backside coatings. The wafers are tested on a cold probe station and then sent out for dicing. LBNL anticipates a delivery rate of 5 wafers/month following an initial 3 month startup phase. DES will occupy about 25% of the production capacity of MSL. The 250 micron thick, fully processed diced CCDs are delivered to FNAL for packaging. Unlike the typical 20 micron thick astronomical CCDs, the LBNL CCDs do not require any processing after packaging.

The CCDs will be packaged at FNAL using a 4-side buttable design similar to that developed by LBNL and the Lick Observatory (Richard Stover). The CCDs are bonded to an Aluminum Nitride board which is a good match in thermal expansion properties to the Silicon. The AlN board contains pads for wirebonding to the CCD readout pads and traces to a connector footprint. Molybdenum has been chosen for the CCD foot material - it is a good match of thermal expansion to the CCDs and has better thermal conductivity than Invar.

The CCDs will be tested and characterized at Fermilab in a new testing facility. We have already begun construction of one test bench. This includes fabrication of a Dewar that can cool and precisely maintain one CCD at a specific temperature (typically 180 K) and the purchase of equipment that can test the CCD optically and with an $Fe^{55}$ source.

The first step in procurement of CCDs for DES was the design of a new 6" wafer mask containing four 2k x 4k devices for the image CCDs, one 2k x 2k device for guide and focus, and eight small devices that fill the remaining space and can be used for initial testing system checkout. The mask used to produce the earlier devices only contained two 2k x 4k devices and was not cost effective for DES. The new DES mask design was completed in March and submitted to DALSA for fabrication. DALSA stipulates a minimum order of a 24 wafers per lot. Fabrication of the first DES wafers began in early April. By mid April a problem with the wafers was discovered (4 of them had large chips on the edges); DALSA suspended processing of this lot and began again. This second lot was completed in mid-May but a higher than normal particulate count was found on the wafers during the final Quality Assurance testing. SNAP has seen similar particulate counts and has found that these devices tend to operate normally, but with an increased number of channel blocks. DALSA has agreed to give this lot of wafers to DES free of charge and to begin a new lot once the contamination has been resolved. These wafers were delivered to LBNL in early June. The thick fully processed devices have been tested on a cold probe station establish that the mask design was successful. Thinning and processing of 5 wafers will take ~12 weeks at LBNL and thus we should have the first thinned fully processed devices by mid-September 2005. We expect that these devices will be sufficient for development of the CCD packaging and testing procedures and will provide valuable experience prior to delivery of the new wafers.

The CCD yield is affected by wafer processing as well as packaging and testing criteria. Based on their experience with a small number of SNAP devices, LBNL has estimated that the CCD yield will be around ~ 25%. We anticipate that LBNL will need to process ~80 wafers (20 in the first lot + 60 in the full order) to obtain 70 science grade devices. Fermilab will use the wafer probing results to decide which CCDs should be packaged. We plan to have parallel CCD packaging and testing equipment and personnel to keep up with the delivery rate from LBNL. DES has adopted a multiple, partially overlapping survey tiling strategy that eases the requirements on the number of bad pixels and columns. As we gain experience with the LBNL CCDs and develop a better understanding of the detailed requirements, we will adjust our device acceptance scheme.

The DES project provides a useful precursor for the SNAP CCD production process. DES will be the first high quantity application for LBNL MSL CCD processing. Success with the DES production will provide a demonstration of capacity for production of SNAP CCDs and is an important milestone towards SNAP itself.

Low noise and reliable CCD readout is essential for obtaining the science results of the DES. The large focal plane naturally leads to long cables and a concomitant risk of electronic interference. The location in the prime focus cage puts severe restrictions on the space and allowed heat dissipation. Although Fermilab has many years of experience building silicon strip detectors for the Tevatron Collider program, there is limited experience reading out CCDs. To get a jump start on learning to read out and test LBNL CCDs we borrowed a CCD readout system from The Amateur Sky Survey (TASS) project. In October 2004, we used this system to read out a standard CCD and by Feb. 2005 we had read out and detected a cosmic ray in an LBNL CCD. In addition, DES has developed a close collaboration with the CCD readout experts at NOAO/Tucson and at CTIO. A week long workshop was held in March in which the various options for DES front end electronics were discussed. This meeting provided face-to-face meetings between the electrical engineers at Fermilab, CTIO, NOAO/Tucson and Barcelona and it built the basis for detailed involvement by all parties in the development of the CCD readout systems for DES.

As a result of the workshop, DES decided to adopt the Monsoon CCD readout system that has been developed by NOAO to capitalize on the effort and experience that has already gone into this design. UIUC has led the effort to acquire the Monsoon hardware and we currently have one system at FNAL and one at UIUC. This system will initially be used for the CCD testing stations. We expect to be reading out an LBNL CCD with Monsoon in the next few weeks. UIUC is also pursuing options for a compact design for the crates; power supplies and their associated thermal control. We have already determined that the density of the existing Monsoon CCD Acquisition board must be increased from 8 to 12 channels to meet the space requirements of DES. Engineers at Fermilab are working with NOAO to rearrange the components and make space for the additional channels. As we gain experience with the Monsoon system in the CCD testing setups, we may determine that modifications are needed to the Clock and Bias board and/or the Master Control board. The group from Barcelona will be directly involved in the development of these boards starting this fall and will ultimately take most of the responsibility for the fabrication of the final components.

The DES 2.2 FOV corrector optics are challenging, but not unprecedented. The design includes large optical elements (the first is 1.1m diameter and has such a high radius of curvature that it must be slumped before figuring) and one aspheric surface. Technology for this design exists but the number of vendors is limited to ~3, and polishing these lenses is a very time consuming process. The Optical Science Laboratory at University College London has extensive experience in the manufacturing and procurement of optical components. We have also recently added groups from the University of Michigan with an experienced optomechanical engineer. The current design of the corrector was developed by Steve Kent and Mike Gladders over a year ago. The UCL and Michigan groups are now reevaluating that design and looking for ways to optimize it for better performance, manufacturability, and cost. Alternative designs will be evaluated and considered at an optical design workshop at Fermilab July 6 and 7.

Detailed DES Timeline:

FY05   R&D  1st CCD run: verify CCD masks (4 2k x 4k devices/wafers)
            Develop CCD testing facility and expertise
            Develop CCD packaging
            Optimize and finalize optical design

FY06 R&D 2nd CCD run: establish firm estimate of the yield
            Optimize and expand CCD testing facilities to deal with production quantitites
            Develop full prototype of Front End electronics and Image collection systems
            Verify slumping of the 1st corrector element maintains optical properties

FY07 Construction
    Production processing of CCD wafers at LBNL
    Production CCD packaging and testing
    Polishing of optical elements
    Production CCD readout and image assembly system
    Camera vessel construction

FY08 Construction
    Complete CCD production, packaging and testing
    Assemble and test full focal plane
    Assemble and test optical corrector

FY09    Assemble and test entire DES Instrument (camera, corrector, electronics, data handling, focus, cooling etc.)

    Ship to Chile and install on Blanco.
    Ready for operation June 09
    First DES observations Oct. 09